\documentclass[preprint]{revtex4}

\usepackage{graphicx}

\def\be{\begin{equation}}
\def\ee{\end{equation}}

\begin{document}

\title{Quantitative Phase Field Model of Alloy Solidification} 

\author{Blas Echebarria\footnote{ Permanent address: Departament de F\'{\i}sica
    Aplicada, Universitat 
Polit\`ecnica de Catalunya, Barcelona, Spain.}}
\affiliation{Physics Department and Center for Interdisciplinary Research on 
Complex Systems, Northeastern University, Boston, Massachusetts 02115;\\
Laboratoire de Physique Statistique, Ecole Normale Sup{\'e}rieure, Paris, France}

\author{Roger Folch}
\affiliation{Instituut-Lorentz, Universiteit Leiden,
Postbus 9506, 2300 RA Leiden, The Netherlands}

\author{Alain Karma}

\affiliation{Physics Department and Center for Interdisciplinary Research on 
Complex
Systems, Northeastern University, Boston, Massachusetts 02115}

\author{Mathis Plapp}

\affiliation{Laboratoire de Physique de la Mati{\`e}re Condens{\'e}e, 
CNRS/Ecole Polytechnique, 91128 Palaiseau, France}

\date{\today}

\begin{abstract}
We present a detailed derivation and thin interface analysis of a  
phase-field model that can accurately simulate 
microstructural pattern formation for low-speed 
directional solidification of a dilute binary alloy.  
This advance with respect to previous phase-field models
is achieved by the addition of a phenomenological 
``antitrapping'' solute current in the mass conservation relation
{\rm [}A. Karma, Phys. Rev. Lett {\bf 87}, 115701 (2001){\rm ]}.
This antitrapping current counterbalances the physical, albeit artificially 
large, solute trapping effect generated when   
a mesoscopic interface thickness is used to simulate 
the interface evolution on experimental length and time scales.
Furthermore, it provides additional freedom in the model
to suppress other spurious effects that scale with   
this thickness when the diffusivity is unequal in solid
and liquid {\rm [} R. F. Almgren, SIAM J. 
Appl. Math {\bf 59}, 2086 (1999){\rm ]}, 
which include surface diffusion and a curvature 
correction to the Stefan condition. 
This freedom can
also be exploited to make the kinetic
undercooling of the interface arbitrarily small even for
mesoscopic values of both the interface thickness and the 
phase-field relaxation time, as  
for the solidification of pure melts 
{\rm [}A. Karma and W.-J. Rappel, Phys. Rev. E {\bf 53}, R3017
(1996){\rm ]}. The performance of the model is demonstrated
by calculating accurately for the first time within a phase-field 
approach the Mullins-Sekerka stability spectrum of a planar interface 
and nonlinear cellular shapes for realistic alloy parameters
and growth conditions.

\end{abstract}

\maketitle

\section{Introduction and summary}
\label{intro}
 
In recent years, the phase-field method has become a standard
tool to simulate microsctructure evolution in 
materials \cite{ARMR}, a subject of both fundamental 
and applied interest \cite{CrossHoh,Kurz}, and more generally to
tackle free boundary problems. Its chief
advantage is to avoid front tracking 
by making phase boundaries spatially diffuse 
with the help of order parameters, termed
phase fields, which vary smoothly between bulk phases.

Simulating the evolution of 
complex morphologies in two and three dimensions 
is in principle straightforward with this method. 
Making quantitative predictions 
on experimentally relevant length and time scales, however,  
has been a major challenge. This challenge stems from the
fact that phase-field simulations are simply not feasible if
parameters of the model are chosen to match those of a real physical
system. With such a choice, both the 
width $W$ of the diffuse interface    
and the characteristic dissipation
time scale $\tau$ in the phase-field kinetics are microscopic:
$W$ is a few Angstroms and $\tau$ is roughly the ratio of $W$ 
and the thermal velocity of atoms in
the liquid \cite{BGJ,MC,HAK}. In contrast, diffusive transport
of solute in bulk phases occurs on macroscopic length and time scales
that are several orders of magnitude larger 
than $W$ and $\tau$, respectively. Therefore,
resolving both microscopic and macroscopic length/time scales in phase-field 
simulations for typical experimental solidification
rates of $\mu$m/sec to mm/sec is impractical, even with
efficient algorithms. 

In view of this, the only possible choice is to carry out simulations
with $W$ and $\tau$ orders of magnitude larger than in a real material.
The question becomes then whether the phase-field model
is still quantitatively meaningful with such a choice. The rest of this
section explores the answer to this question in the context of
previous works and serves both as a summary and a guide
for the following sections of this paper. 
To conclude this section, we summarize
the mains results needed to carry out quantitative simulations of
the directional solidification of a dilute binary alloy. 

\subsection{Capillarity} 
 
In the phase-field model 
of a pure substance (of say $A$ molecules), 
the excess free-energy of
the solid-liquid interface, $\gamma$, is determined by
the combination of the bulk free-energy
density at the melting point, $f(\phi,T_m)$, 
which is a double-well function
with minima corresponding to solid 
and liquid and the gradient square term, 
$\sigma |\vec\nabla \phi|^2$. Minimization
of the total free-energy, which is the spatial integral of the sum
of these two terms, yields the standard result that
$\gamma\sim WH$, where $H$ is the barrier height of the 
double-well potential, and $W\sim (\sigma/H)^{1/2}$ is the
width of the $\phi$ tanh-profile in the diffuse interface.
This results implies that there always exist 
a pair of values of $\sigma$ and $H$ for
any pair of values of $W$ and $\gamma$.
Thus, the experimental magnitude of 
$\gamma$ in the classic Gibbs-Thomson
condition can be reproduced even if   
a computationally tractable ``mesoscopic" interface thickness (i.e. on
a scale comparable to the microstructure) is used in the phase-field model.
Optimally, this thickness should be chosen just small enough to resolve 
accurately the interface curvature. 

A phase-field model for a dilute alloy can 
generally be constructed by adding
to the free-energy density the 
contribution of solute $B$ molecules.
The simplest way to construct this free-energy is to interpolate 
between the known free-energy densities 
in solid and liquid with a single
function of $\phi$, as in the original model of
Wheeler {\it et al.} \cite{Wheeler} (see also Ref. \cite{Caginalp93}). 
From a computational standpoint, however, 
this approach places a stringent constraint
on the interface thickness. The reason is that  
there is generally an extra contribution to $\gamma$ due to 
solute addition that depends on interface thickness,
solute concentration at the interface, and temperature.  
In section III.A, we show how this extra contribution 
can generally be made to vanish by using
two different functions of $\phi$,  
which interpolate separately between solid and liquid
the enthalpic (internal energy)
and entropic part of the free-energy density. The condition that
this contribution vanishes takes the form of an algebraic relation
between these two interpolation functions. If this relation is satisfied,
the model introduced previously in Ref. \cite{encyclo} is recovered.
The equilibrium phase-field profile decouples from the equilibrium solute 
concentration profile and $\gamma\sim WH$, as for a pure substance. 
This removes the constraint on the interface thickness associated with solute 
addition without the need to introduce separate concentration fields in
each phase as in Refs. \cite{Steinbach98,Kim99}.
 
\subsection{Interface-thickness-dependent nonequilibrium effects}

The main conclusion from the preceding paragraphs is that the
phase-field method provides sufficient freedom to choose $W$ 
arbitrarily large to model capillarity.
However, microstructural pattern formation is also generally controlled
by nonequilibrium effects at the interface. For a microscopic $W$
and low solidification velocities, these 
effects are negligibly small. The interface
relaxes rapidly to a local thermodynamic equilibrium and 
its nonlinear evolution is driven by slowly evolving  
gradients of thermodynamic quantities in bulk phases.
For a mesoscopic thickness, however, these nonequilibrium effects 
become artificially magnified, thereby competing
with, or even superseding, capillary effects, and  
dramatically altering the large scale pattern evolution. Therefore, the 
central challenge of quantitative phase-field modeling of solidification
at low velocity, onto which we focus in the present work, consists of formulating
the model, and knowing how to choose its parameters, in order to avoid 
unphysically large non-equilibrium effects at the interface. This is in contrast
to rapid solidification where
nonequilibrium effects play a dominant role. 
In this case, the challenge consists
of describing the correct magnitude of these effects with mesoscale
phase-field parameters, which
requires a different approach (see Ref. \cite{Bragard}).

For pure materials, Karma and Rappel \cite{KarRap}
have developed a thin interface analysis, which only assumes that   
$W$ is small compared 
to the scale of the microstructure. This analysis 
shows that the standard free-boundary problem 
of solidification $-$ a classic Stefan condition together
with a velocity-dependent form of the Gibbs-Thomson relation 
that incorporates interface kinetics $-$ 
is recovered even for a mesoscopic $W$.
Heat diffusion in the mesoscale interface region only modifies
the expression for the interface kinetic coefficient, $\mu_k$. 
This ``renormalization'' of $\mu_k$ has the crucial property 
that $\tau$ needs not be microscopic
to make this coefficient arbitrarily large (arbitrarily fast
kinetics), and hence to simulate the limit of local equilibrium 
at the interface dominated by capillarity.

This advance bridges the gap between the atomistic scale of interfacial
phenomena and the mesoscale of the microstructure.
In addition, efficient multi-scale simulation 
algorithms have been developed to bridge the remaining gap between 
the microstructure and the transport scales \cite{Proetal,PlaKar}.
The combination of these two advances has lead to the first
direct quantitative comparison between 
fully three-dimensional phase-field simulations of dendritic
growth in pure melts at low undercooling and experiments \cite{KarTip}.

Achieving the same success for alloys has turned out to be
considerably more difficult. A major source of difficulty is 
that solute diffusion is generally much slower in solid
than liquid. When diffusion is asymmetrical, 
the use of a mesoscopic $W$ artificially magnifies several nonequilibrium
effects at the interface that are absent when diffusion is symmetrical.
Consequently, phase-field models in which one or several of these effects
are present \cite{Wheeler,Caginalp93,Warren95,Steinbach98,Kim99}
are not suitable for quantitative simulations at low velocity.

These nonequilibrium effects were first characterized in
detail by Almgren \cite{Alm} using a thin 
interface analysis of a phase-field model of the 
solidification of pure melts with asymmetric diffusion. 
Directly analogous effects are present in 
alloy solidification \cite{KarmaPRL}, 
which include (i) solute diffusion  
along the arclength of the interface (surface diffusion), (ii)
a modification of mass conservation associated with 
the local increase of arclength of a moving curved interface
(interface stretching), and (iii) a discontinuity
of the chemical potential of the dilute impurity 
across the interface. 

These nonequilibrium effects originate physically
from solute transport in the mesoscale interface region
that is governed by the standard continuity 
equation for a dilute alloy
\begin{equation}
\frac{\partial c}{\partial t}=\frac{Dv_0}{RT_m}
\vec \nabla \cdot \left( \tilde q(\phi) c \vec \nabla  \mu \right),\label{MASS}
\end{equation} 
where $R$ is the gas constant, $v_0$ is the molar volume
of solute molecules, $T_m$ is the melting temperature,
$\mu$ is the chemical potential, and the product $D\tilde q(\phi)$ 
governs how the solute diffusivity varies through the 
diffuse interface, from zero in the solid (for a one-sided
model) to a constant value $D$ in the liquid.
The best known of these effects 
is solute trapping \cite{KurFis,trapping} that is associated
with the chemical potential jump at the interface. 
The problem is that the magnitude 
of all these effects scales with the interface thickness. 
Since $W$ in phase-field computations is 
orders of magnitude larger than in reality, solute trapping will
become important at growth speeds where it is 
completely negligible in a real material. Surface
diffusion and interface stretching, in turn, modify the 
mass conservation condition
\begin{equation}
c_l(1-k)V_n=-D\frac{\partial c}{\partial n}+\cdots\label{stef}
\end{equation}
where $c_l$ is the concentration on the liquid side of the
interface, $k$ is the partition coefficient, $V_n$ is the
normal interface velocity, and $``\cdots"$
is the sum of a correction $\sim c_l(1-k)WV_n{\cal K}$, corresponding
to interface stretching, where ${\cal K}$ is the local interface
curvature, a correction  $\sim WD\partial^2 c_l/\partial s^2$, 
corresponding to surface diffusion along the arclength 
$s$ of the interface, and a correction $\sim kc_l(1-k)WV_n^2/D$ proportional
to the chemical potential jump at the interface. 
All three corrections, which are proportional to the interface 
thickness, are negligible in a real material at low velocity.
For this reason, they have not been traditionally considered in   
sharp-interface models (reviewed in Sec. II). 
For a mesoscopic interface thickness, however,
the magnitude of these corrections becomes comparable
to the magnitude of the normal gradient of solute, thereby modifying
$V_n$ and the pattern evolution. Thus, the
phase-field model must be formulated to make all three
effects vanish. 

\subsection{Limitation of variational models}

The model discussed in Sec. III.A follows the general approach
of nonequilibrium thermodynamics where the evolution equations  
for $\phi$ and $c$ are derived variationally from a
Lyapounov functional $\cal F$ that represents the total free-energy of
the system. The resulting ``gradient dynamics'' 
guarantees that $\cal F$ decreases monotonously 
in time in an isolated system. In addition to the double-well
potential $f(\phi)$, this variational model contains 
three basic interpolation functions: the two
functions that interpolate between solid and
liquid the enthalpic and entropic part of the free-energy density
(section III.A), which we denote here by $\bar g(\phi)$ and $\tilde g(\phi)$,
respectively, and the diffusivity function $\tilde q(\phi)$
in Eq.~(\ref{MASS}) that varies from zero in solid to unity in liquid.

These functions should in principle be chosen to cancel all
spurious interface-thickness dependent effects. 
As already discussed in Sec. I.A, a quantitative
description of capillarity  
can be obtained by requiring that the solute contribution to $\gamma$
vanish. This condition is only satisfied if the two functions
$\bar g(\phi)$ and $\tilde g(\phi)$ are related, and the latter
determines the equilibrium solute concentration profile   
\begin{equation}
c_0(\phi)=\frac{c_s+c_l}{2}+\tilde g(\phi)\frac{c_s-c_l}{2},\label{C0}
\end{equation} 
where $\tilde g(\phi)$ varies from $+1$ in the solid where $c_0=c_s$
to $-1$ in the liquid where $c_0=c_l$. 

We are left with only
two functions, $\tilde g(\phi)$ and $\tilde q(\phi)$, to satisfy 
the three aforementioned conditions that surface diffusion, 
interface stretching, and the chemical
potential jump at the interface, should vanish. The 
thin-interface analysis of section IV applied to this variational model
shows that these three conditions are given, respectively, by 
\begin{eqnarray}
\int_{-\infty}^0 
dr  \, \tilde q(\phi(r)) c_0(\phi(r)) 
&=&
\int_0^{+\infty} 
dr \left[c_l-\tilde q(\phi(r)) c_0(\phi(r))\right],  
\label{SD}\\
\int_{-\infty}^0 
dr \left[c_0(\phi(r))-c_s\right] 
&=&
\int_0^{+\infty} 
dr \left[c_l-c_0(\phi(r))\right],  
\label{STR} \\
\int_{-\infty}^0 
dr \frac{c_0(\phi(r))-c_s}
{\tilde q(\phi(r)) c_0(\phi(r))} 
&=&
\int_0^{+\infty} 
dr \left[(1-k)-\frac{c_0(\phi(r))-c_s}
{\tilde q(\phi(r)) c_0(\phi(r))}   \right]
\label{MU}
\end{eqnarray}
where $k\equiv c_s/c_l$ is the partition coefficient,
$r$ is the coordinate normal to the solid-liquid interface
that varies from $-\infty$ in solid to $+\infty$ in liquid far
from the interface, and $c_0$ is given by Eq. (\ref{C0}) 
that can be assumed to remain valid for 
a slowly moving interface.  

\begin{figure}[ht]
\includegraphics[width=7cm]{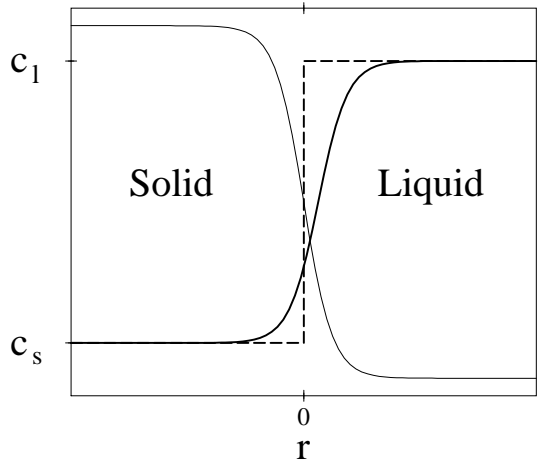}
\caption{Illustration of the definition of surface excess.
The excess of solute is the integral along $r$ of 
the actual solute profile (thick solid line) minus
its step profile idealization (thick dashed line) with the 
Gibbs dividing surface at $r=0$. This excess is negative
in the depicted example. The thin solid line 
depicts the phase-field
profile, $\phi(r)=-\tanh(r/\sqrt{2}W)$. 
The standard mass conservation
condition [Eq.~\protect(\ref{stef})] is recovered if all three
excess quantities defined by the difference between the
left-hand-side and the right-hand-side of 
Eqs.~\protect(\ref{SD})-(\ref{MU}), vanish.}
\label{gibbs}
\end{figure}

A simple physical interpretation 
of these conditions is obtained by analogy with Gibbs' treatment 
of interfacial phenomena where ``excess quantities''
are attributed to a mathematical surface with zero volume
dividing two phases, which corresponds here to $r=0$. 
In this analogy, Eqs.~(\ref{SD})-(\ref{MU}) are
conditions that excess quantities of the interface vanish.
For example, as illustrated in Fig.~\ref{gibbs},
the excess of solute is the integral through
the diffuse interface of the difference between 
the actual smoothly varying solute profile $c_0$ and the imaginary
step function profile equal to $c_s$ for $r<0$ and $c_l$ for
$r > 0$. The condition that this excess vanishes is identical
to Eq.~(\ref{STR}). It implies that mass conservation is left
unchanged if there is no excess of solute to redistribute 
along the arclength of the interface.
Similarly, surface diffusion vanishes [Eq.~(\ref{SD})]
if there is no excess of the transport coefficient 
$\tilde q(\phi) c$ multiplying the chemical potential
gradient in Eq.~(\ref{MASS}). Finally, the jump of
chemical potential vanishes if there is no excess of chemical 
potential gradient [Eq.~(\ref{MU})]. This condition is simple to derive for
a flat interface by rewriting Eq.~(\ref{MASS})
in a local frame moving at velocity $V$ (i.e. $\partial/\partial t
\rightarrow -V\partial/\partial r$ and $\vec \nabla \rightarrow 
\partial/\partial r$). After integrating both sides of Eq. 
(\ref{MASS}) once with respect to $r$, 
one obtains the expression for the chemical potential
gradient through the diffuse interface
$\partial \mu/\partial r \approx 
-V(c_0-c_s)RT_m/(Dv_0\tilde qc_0)$, and hence Eq.~(\ref{MU}). 

A major pitfall of this variational model is that all three
excess quantities cannot be made to vanish simultaneously with
only two functions $\tilde g(\phi)$ and $\tilde q(\phi)$.
For example, with the standard quartic 
form of the double-well, which is an even function of $\phi$, the
equilibrium $\phi$ profile is an odd function of $r$.
Therefore, Eq.~(\ref{STR}) can be made to vanish by choosing
$\tilde g(\phi)$ to be an odd function of $\phi$. It is
then possible to choose $\tilde q(\phi)$ to satisfy Eq.~(\ref{SD}).
However, this leaves no freedom to make
the jump of chemical potential vanish. 
More generally,
it is possible to make two of the three excess quantities
vanish for different choices of $\tilde g(\phi)$ and $\tilde q(\phi)$, 
but not the three of them simultaneously. 

Elder {\it et al.} \cite{Eldetal01}
proposed to make the discontinuity of chemical
potential vanish by an appropriate choice of interface position
(Gibbs dividing surface) which makes the corresponding excess
quantity vanish. These authors, however, did not take into
account the other two excess quantities found
by Almgren for asymmetric diffusion \cite{Alm}. These quantities appear
at higher orders in the asymptotic expansion used by 
Elder {\it et al.} which, for the 
solidification of pure melts with
symmetrical diffusion, yields the same 
results as the thin interface analysis
of Karma and Rappel \cite{KarRap}. For asymmetrical diffusion, all three 
excess quantities can generally not be made 
to vanish by a redefinition of the interface position.

It might be possible to make all 
three excess quantities vanish for
non-trivial oscillatory forms of the functions
$\tilde g(\phi)$ and $\tilde q(\phi)$.  
Such forms, if they exist, would require a high resolution of the
interfacial layer that is not computationally desired.
Also, other variational models than the one discussed here
are in principle possible.  
McFadden {\it et al.} have formulated a variational phase-field model of the 
solidification of pure melts with unequal thermal conductivities
\cite{McFadden00}. This model provides 
additional freedom to cancel the discontinuity of temperature
at the interface, which they interpret as ``heat trapping'' by analogy with 
solute trapping that is associated with the discontinuity of 
chemical potential in the case of alloys. However, as Elder {\it et al.},
these authors did not consider the additional constraints associated
with surface diffusion and interface 
stretching for a non-planar interface. While we cannot rule out
that it may be possible to formulate variational models that remove all   
constraints on the interface thickness, achieving 
this goal appears extremely difficult.
 
\subsection{Non-variational models and antitrapping}

A way out of this impasse is to drop the requirement that
the equations of the phase-field model be strictly variational.
This provides additional freedom to cancel all spurious
corrections produced by a mesoscale interface thickness.
As shown recently in Ref. \cite{KarmaPRL},  
a successful approach consists of
adding a phenomenological ``antitrapping current''  
in the continuity relation [Eq. (\ref{MASS})]. This current
produces a net solute flux from solid to liquid proportional
to the interface velocity that counteracts solute trapping and
restores chemical equilibrium at the interface.
By adjusting the magnitude of this current,
which modifies Eq. (\ref{MU}), it is therefore
possible to satisfy simultaneously Eqs. (\ref{SD})-(\ref{MU}).

Furthermore, the same function 
$\tilde g(\phi)$ must appear in the evolution equations for $\phi$
and the continuity relation [Eq. (\ref{MASS})] 
in the variational model. The additional freedom to replace 
$\tilde g(\phi)$ by another function $h(\phi)$ in the modified
continuity relation with the antitrapping current turns out 
to be critically important to obtain the same
renormalization of the interface kinetic coefficient
as in the analysis of Karma and Rappel 
for pure melts \cite{KarRap}. 

\subsection{Summary of phase-field equations and thin-interface limit}

We summarize here the equations of 
the non-variational phase-field model
for the directional solidification of a 
dilute binary alloy that are needed to carry 
out quantitative computations. The lengthy
details of the derivation of the model and of 
the asymptotic analysis are exposed in 
sections III and IV below. The model uses the
standard low velocity frozen temperature
approximation, $T=T_0+G(z-V_pt)$, where $V_p$ is the
pulling speed and $G$ is the temperature gradient. 
The basic equations of the model are
\begin{eqnarray}
& &\tau(T)\frac{\partial \phi}{\partial t} = W^2\nabla^2\phi
+\phi-\phi^3-\frac{\tilde \lambda}{1-k}\tilde g'(\phi)
\left(e^u - 1 - \frac{T-T_0}{mc_l^0} \right), \\
& &{\partial c\over \partial t} = 
    \vec{\nabla}\left(D\tilde q(\phi)c \vec{\nabla} u - j_{at}\right),
\end{eqnarray}
where
\be
u=\frac{v_0}{RT_m}(\mu-\mu_E)=\ln\left(2c\over c_l^0[1+k-(1-k)h(\phi)]\right)
\ee
is a dimensionless measure of the deviation of the 
chemical potential from its equilibrium value $\mu_E$ at a
reference temperature $T_0$ with corresponding liquidus
concentration $c_l^0$, $m<0$ is the liquidus slope,
\be
j_{at} =  
  - a W (1-k) c_l^0 e^u 
    \frac{\partial\phi}{\partial t} {\vec \nabla \phi\over |\vec\nabla\phi|},
\ee
is the antitrapping current,
\be
\tau(T)=\tau_0\left(1+\frac{T-T_0}{mc_l^0}\right)
\ee
is a temperature-dependent phase-field relaxation time,
and $\tilde \lambda$ is a dimensionless coupling constant. 
For the choices $h(\phi)=\phi$, $\tilde q(\phi)=(1-\phi)/[1+k-(1-k)\phi]$, 
$\tilde g(\phi)=(15/8)(\phi-2\phi^3/3+\phi^5/5)$, and $a=1/(2\sqrt{2})$,
this model reduces in its thin-interface limit to the standard one-sided
model of alloy solidification. The chemical
capillary length $d_0$ and the interface kinetic coefficient 
$\beta$ (defined in section II) are related to 
the phase-field parameters by  
\begin{eqnarray}
d_0&=&a_1W/\lambda,\label{param1}\\
\beta&=&a_1 \frac{\tau(T)}{\lambda W}\left[
1-a_2\frac{\lambda W^2}{\tau_0 D}\right],\label{param2}
\end{eqnarray}
where $\lambda=15\tilde\lambda/8$ and $a_1=5\sqrt{2}/8$ 
and $a_2=0.6267$ are the same numerical constants 
as in Ref. \cite{KarRap}; we note that $\tilde \lambda$ has
been defined for convenience in the present paper 
to avoid carrying a numerical factor of $15/8$
in the thin-interface analysis of the equations.
 
A previous version of this
model for isothermal alloy solidification 
was presented in Ref. \cite{KarmaPRL} 
together with benchmark computations for dendrite growth.
The present extension to non-isothermal
growth conditions introduces a temperature-dependent
relaxation time $\tau(T)$. As discussed
in more details in section IV.C, 
this new feature makes it possible
to achieve vanishing interface kinetics (i.e. local
equilibrium at the interface) for the entire
range of interface temperature that occurs during
directional solidification. For simplicity, we have
written down the equations of the model for isotropic
surface tension and interface kinetics. The extension
to anisotropic growth is discussed in section IV.E.
Also, both for simulating and analyzing the above equations,
it is convenient to rewrite them in terms of a new
variable $U=(e^u-1)/(1-k)$. This avoids numerical
computations of exponential and logarithm functions.
In addition, it transforms the equations in a form closely related
to the phase-field model for the solidification of 
a pure substance where $U$ is the direct analog of
the temperature field. Details of this change of 
variable are given in section III.B.c. 

Simulations of microstructural pattern formation
using this model are presented in section V, which
also contains the final form of the anisotropic
phase equations (\ref{pfn1}) and (\ref{pfn2}) that
are solved numerically.
We report 
the first ever quantitative phase-field computation 
of the classic Mullins-Sekerka linear stability spectrum of
a planar interface \cite{MS} and nonlinear cell shapes  
for realistic experimental parameters of low velocity
directional solidification.

\section{Sharp-interface models}
\label{sharp}

We consider the solidification of a dilute binary alloy made
of substances A and B, with an idealized phase diagram that 
consists of straight liquidus and solidus lines of slopes 
$m$ and $m/k$, respectively, where $k$ is the partition 
coefficient. The interface is supposed to be in local 
equilibrium, that is,
\be
c_s = k c_l,
\label{partition}
\ee
where $c_s$ and $c_l$ are the concentrations (in molar fractions)
of impurities B at the solid and liquid side of the
interface, respectively. 

The interface temperature satisfies
the generalized Gibbs-Thomson relation,
\be
T = T_m - |m| c_l - \Gamma {\cal K} - V_n/\mu_k,
\label{githo}
\ee
where $T_m$ is the melting temperature of pure $A$, 
\be
\Gamma = \frac{\gamma T_m}{ L},
\label{githoconst}
\ee
the Gibbs-Thomson constant, $\gamma$, the 
surface tension, $L$, the latent heat of fusion per
unit volume, ${\cal K}$, the interface curvature, $V_n$ its
normal velocity, and $\mu_k$ the linear kinetic coefficient. 
Here, the surface tension and the kinetic coefficient are taken isotropic 
for simplicity; anisotropic interface properties will be considered 
below. 

Heat is supposed to diffuse much faster than impurities, 
so that the temperature field can be taken
as fixed by external conditions, in spite of the rejection of
latent heat during solidification. 
Then, Eq.~(\ref{githo}) yields a boundary 
condition for the solute concentration at the interface.

Of particular interest is the {\em one-sided} model of
solidification that assumes zero diffusivity in the solid. 
This is often a good approximation for alloy solidification, 
in which the solute diffusivity in the solid may be several 
orders of magnitude lower than in the liquid.

\subsection{Isothermal solidification}
\label{isothermal}

For isothermal solidification at a fixed temperature $T_0<T_m$,
the concentration obeys the set of sharp-interface equations
\begin{eqnarray}
\partial_t c&=&D\nabla^2c,\label{fb1} \\
c_l(1-k)V_n&=&-D\partial_nc|^+,\label{fb2}\\
c_l/c_l^0&=&1-(1-k)d_0\,{\cal K}-(1-k)\beta V_n\label{fb3}
\end{eqnarray}
where $D$ is the solute diffusivity in the liquid,
$V_n$, the normal velocity of the interface, $\partial_nc|^+$,
the derivative of the concentration field normal to 
the interface, taken on the liquid side of the interface,  
\be
\label{cref}
c_l^0 = (T_m-T_0)/|m|,
\ee
the equilibrium concentration 
of the liquid at $T_0$, 
\be
d_0={\Gamma \over \Delta T_0},
\label{d0def}
\ee
the chemical capillary length, 
where $\Delta T_0=|m|(1-k)c_l^0$ is the freezing range,
and $\beta=1/[\mu_k \Delta T_0]$. 
Equation (\ref{fb2}), the Stefan condition, expresses
mass conservation; Eq.~(\ref{fb3}) can be directly
obtained from Eq.~(\ref{githo}).

\subsection{Directional solidification}
\label{directional}

For directional solidification, we use
the frozen temperature approximation, in which the 
temperature field for solidification with speed
$V_p$ in a temperature gradient of magnitude $G$ 
directed along the $z$ axis is taken as
\be
T(z) = T_0 + G(z-V_pt).
\label{tempfield}
\ee
Now $T_0$ is given by inverting Eq. (\ref{cref}),
and $c_l^0 = c_\infty/k$, where
$c_\infty\equiv c(z=+\infty)$ is the global sample 
composition. Thus, $c_l^0$ is the solute concentration
on the liquid side of a steady-state planar interface.
Then, Eq.~(\ref{fb3}) is replaced by
\be
c_l/c_l^0 = 1-(1-k)d_0\,{\cal K} -(1-k)\beta V_n -(1-k)(z-V_pt)/l_T
\label{fb3direc}
\ee
where
\be
l_T={ |m|(1-k) c_l^0\over G}
\ee
is the thermal length.

\subsection{Formulation in terms of dimensionless supersaturation}
\label{uparamsharp}
In order to later compare with
the sharp-interface limit of the phase-field models treated here,
we rewrite Eqs. (\ref{fb1},\ref{fb2},\ref{fb3direc})
in terms of the local supersaturation with respect to
the point $(c_l^0,T_0)$, measured
in units of the equilibrium concentration gap
at that point,
\be
\label{udefsharp}
U=\frac{c-c_l^0}{c_l^0(1-k)}.
\ee
We obtain
\begin{eqnarray}
\label{fbU1}
\partial_t U & = & D\nabla^2 U \quad{\rm (liquid)},  \\
\label{fbU2}
[1+(1-k)U]V_n & = & -D \partial_n U|^+ \quad{\rm (interface)},  \\
\label{fbU3}
U & = & -d_0 {\cal K} -\beta V_n - (z-V_p t)/l_T \quad{\rm (interface)”}.
\end{eqnarray}
Note that, for $k=1$, we recover the constant miscibility gap model.
Furthermore, if we reinterpret $U$ as a dimensionless temperature
and drop the directional solidification term $(z-V_p t)/l_T$,  
we obtain a one-sided version of the pure substance model.

\section{Phase-field models}
\label{diffuse}
In this section we first derive a generic variational model 
(Sec. \ref{variational}), and we then modify it in view of canceling 
spurious effects (Sec. \ref{nonvariational}).

\subsection{Variational formulations}
\label{variational}

In a phase-field model, a continuous scalar field $\phi$
is introduced to distinguish between solid ($\phi=+1$) and liquid
($\phi=-1$).
The two-phase system is usually described by a phenomenological
free energy functional,
\be
F[\phi,c,T] = \int_{dV} \left[\frac{\sigma}{2}|\vec \nabla \phi|^2
+f(\phi,T_m)
+f_{AB}(\phi,c,T)\right],
\label{fint0}
\ee
where 
\begin{equation}
f(\phi,T_m)=H(-\phi^2/2+\phi^4/4) \label{dw}
\end{equation}
is the standard form of a double-well potential 
providing the stability of the two phases $\phi=\pm 1$
with a barrier height $H$, 
$f_{AB}(\phi,c,T)$ changes their relative stability as a function
of the position in a $T$--$c$ phase diagram, and the term in
$\sigma$ provides a penalty for phase gradients which ensures
a finite interface thickness. $H$ has dimensions of energy per unit
volume, and $\sigma$ of energy per unit length.
 
In a variational formulation, the equations of motion for all
fields (here the concentration and phase fields) can be derived
from that functional:
\begin{eqnarray}
\frac{\partial \phi}{\partial t} &=&
-K_\phi\frac{\delta F}{\delta \phi}
\label{pa1},\\
\frac{\partial c}{\partial t}&=&\vec \nabla \cdot \left(M(\phi,c)
\vec\nabla \, \frac{\delta F}{\delta c}\right),
\label{pa2}
\end{eqnarray}
where $K_\phi(T)$ is a kinetic constant that can generally
be temperature-dependent. The second equation is 
a statement of mass conservation, since it can be rewritten as
\begin{equation}
\frac{\partial c}{\partial t}+\vec\nabla \cdot\vec J_c\,=\,0,
\end{equation}
where $\vec J_c=-M\vec\nabla \mu$ is the solute current density,
$\mu\equiv \delta F/\delta c$ is the chemical potential,
and $M(\phi,c)$ is the mobility of solute atoms or molecules,
which we choose to be
\be
M(\phi,c)= {v_0\over RT_m} D \tilde q(\phi) c
\ee
in order to later obtain Fick's law of diffusion in the liquid.
Here, $v_0$ is the molar volume of A, $R$, the gas constant, and
$\tilde q(\phi)$, a dimensionless function
that interpolates between 0 in the solid and 1 in the liquid,
and hence dictates how the solute diffusivity varies through
the diffuse interface.
Note that we have not included an equation of motion for
the temperature field, since we consider it fixed by external
constraints. Of course, the formalism could be extended to
include an appropriate equation for heat transfer \cite{Wangetal}.

An important step is the construction of the function
$f_{AB}$ that interpolates between the free energy densities
of the bulk phases (solid and liquid). 
While these bulk free energies should reduce to the
curves that can be obtained from thermodynamic databases,
the dependence of $f_{AB}$ on $\phi$ influences only
the interfacial region, and this freedom can be used to construct a
particularly simple phase-field model. This will be illustrated
here for the case of a dilute binary alloy.
First, we consider the bulk free energies and make sure that they
reproduce the equilibrium properties of the sharp-interface model
of Sec. \ref{sharp}. Then, we interpolate between them.

For a dilute alloy, the free energies
of solid and liquid $f_\nu (c,T)$ can be written as the sum of the 
free energy of pure A, $f_\nu^A(T)$, 
and contributions due to solute addition:
\begin{equation}
f_\nu(c,T) = f_\nu^A(T) + {RT\over v_0} (c \ln c - c) + 
   \varepsilon_\nu c \quad \nu = {\rm l, s}.
\end{equation}
The second term on the right hand side is the dilute form of the
mixing entropy, and the term $\varepsilon_\nu c$ is the change of the 
internal energy density. 
We expand this
expression to first order in $T-T_m$ to recover the
straight liquidus and solidus lines of Sec. \ref{sharp}: 
\begin{equation}
f_\nu(c,T) = f_\nu^A(T_m) - s_\nu (T-T_m) 
+ {RT_m\over v_0} (c \ln c - c) + 
\varepsilon_\nu c ,
\end{equation}
where $s_\nu = -\partial f_\nu^A/\partial T$ are the entropy densities
of solid and liquid at $T_m$. By using $T_m$ instead of $T$ in 
the mixing entropy, we have neglected terms of order $(T-T_m)c$, 
which are second-order for dilute alloys.

The phase diagram is determined by the standard
common tangent construction, which is equivalent to requiring 
that the chemical potential and the grand potential $\omega$
(i.e., the thermodynamic potential for a varying number of solute 
particles) be equal in the solid and liquid. The corresponding
equilibrium concentrations $c_s(T)$ and $c_l(T)$ are
the solutions of
\begin{eqnarray}
\left. \frac{\partial f_s(c,T)}{\partial c}\right|_{c=c_s} & = &
\left. \frac{\partial f_l(c,T)}{\partial c} \right|_{c=c_l}
=\mu_E(T), \label{ct1}\\
f_s(c_s,T)-\mu_E \,c_s & = &
 f_l(c_l,T)-\mu_E\, c_l = \omega_E(T).\label{ct2}
\end{eqnarray}

The first equality yields the partition relation Eq.~(\ref{partition}),
$c_s=kc_l$, with a partition coefficient
\be
k = \exp\left(-{v_0 \Delta \varepsilon\over RT_m}\right),
\label{kdef}
\ee
where we have defined $\Delta\varepsilon = \varepsilon_s - \varepsilon_l$.
Combining this result with Eq.~(\ref{ct2}) yields
\be
c_l = {Lv_0 \over T_m^2 R (1-k)} (T_m-T),
\label{liquidconc}
\ee
where we have used that the latent heat per unit volume is
$L=T_m(s_l-s_s)$. From Eq. (\ref{liquidconc}), 
we identify the liquidus slope to be
\be
m = - {T_m^2 R (1-k)\over v_0 L},
\label{clauclap}
\ee
the van't Hoff relation for dilute binary alloys.

In the standard phase-field approach, the two bulk free energies are 
interpolated with the help of a single function of the 
phase field $\phi$. Here, it is advantageous to use two different
interpolation functions for the entropy
and the internal energy terms,
\begin{equation}
f_{AB}(\phi,c,T) = f^A(T_m)-(T-T_m)s(\phi) 
+ {RT_m\over v_0} (c\ln c - c) 
+\varepsilon(\phi)c,
\end{equation}
with
\begin{eqnarray}
\varepsilon(\phi) & = & 
\bar\varepsilon + \bar g(\phi) \Delta\varepsilon/2, \label{einterpolant}\\
s(\phi) & = &
\frac{s_s+s_l}{2} - \tilde g(\phi) \frac{L}{2T_m}, \label{sinterpolant}
\end{eqnarray}
where $\bar \varepsilon = (\varepsilon_s+\varepsilon_l)/2$, and
we have again used $L=T_m(s_l-s_s)$ in $s(\phi)$.
$\tilde g (\pm 1) = \bar g (\pm 1) = \pm 1$, and we further require
$\tilde g'(\pm 1)=\bar g'(\pm 1) = 0$ for $\phi=\pm 1$ to remain
bulk equilibrium solutions for any value of $c$ and $T$.

This completes the model specification, except for the interpolation
functions $\tilde{g}(\phi)$ and $\bar g(\phi)$. In order to choose
them appropriately, it is important to consider the
equilibrium properties of the model, which follow from
the conditions
\begin{eqnarray}
\frac{\delta F}{\delta c}&=&\mu_E,\label{eq1}\\
\frac{\delta F}{\delta \phi}&=&0,\label{eq2}
\end{eqnarray}
where $\mu_E$ is the spatially uniform equilibrium value of the
chemical potential. These two equations uniquely determine the
spatially varying stationary profiles of $c$ and $\phi$ in the diffuse
interface region, $c_0(x)$ and $\phi_0(x)$. Since the phase field
interpolates between the two bulk free energies, the limiting values of
the concentrations and the equilibrium chemical potential are the
ones determined by the common tangent construction above.
From Eq.~(\ref{eq1}), we have
\be
\frac{RT_m}{v_0}\,\ln c_0 \,+\,\bar \varepsilon\,+\,\bar g(\phi_0)
 \frac{\Delta \varepsilon}{2}
=\mu_E, \label{mueq}
\ee
from which we obtain the expression for the equilibrium concentration profile 
using the solution of Eq. (\ref{ct1}) and Eq. (\ref{kdef}),
\begin{equation}
c_0(x)=c_l \exp\left(\frac{\ln k}{2}\left[1+\bar g(\phi_0(x))
\right] \right)=c_l k^{[1+\bar g(\phi_0(x))]/2}.\label{ceq}
\end{equation}
From the equilibrium condition for $\phi$, Eq.~(\ref{eq2}), we obtain
\begin{equation}
 \sigma \frac{d^2\phi_0}{dx^2}+H(\phi_0-\phi_0^3)= {\tilde g'(\phi_0)\over 2}
{T-T_m\over T_m} L + {\bar g'(\phi_0)\over 2} \Delta \varepsilon\, c_0 .
\end{equation}
With the help of Eqs.~(\ref{kdef}), (\ref{liquidconc}) and (\ref{clauclap}), 
the right-hand side can be rewritten as
\begin{equation}
 \sigma \frac{d^2\phi_0}{dx^2}+H(\phi_0-\phi_0^3)= -{RT_m(T-T_m)\over 2v_0m}
   \left[(1-k){\tilde g'(\phi_0)} + \ln k {c_0(x)\over c_l}
       {\bar g'(\phi_0)} \right] .
\label{phieq}
\end{equation}
For a generic choice of the functions $\tilde g$ and $\bar g$, and
in particular for the ``standard'' choice $\tilde g = \bar g$, 
no analytic solution for $\phi$ is known. Furthermore, the equilibrium
solution and its properties, in particular its surface tension,
depend on the various coefficients that appear in the right-hand
side. This can be avoided if the right-hand side vanishes 
($\partial_\phi f_{AB}(\phi_0,c_0,T)=0$). With the help of 
Eq.~(\ref{ceq}), we obtain the condition on the interpolation functions,
\be
(1-k){\tilde g'(\phi)\over 2} + \ln k {\bar g'(\phi)\over 2} 
    \exp\left(\frac{\ln k}{2}\left[1+\bar g(\phi)\right] \right) = 0.
\label{gcond}
\ee
It can be used to eliminate one of them in terms of the other. 
Taking into account the requirement 
$\tilde g(\pm 1) = \bar g(\pm 1) = \pm 1$, we find
\be
\tilde g(\phi) = 
  {1+k - 2\exp\left(\frac{\ln k}{2}\left[1+\bar g(\phi)\right]\right)\over 1-k}
  = {1+k - 2k^{[1+\bar g(\phi)]/2}\over 1-k} ,
\label{gsol}
\ee
\be
\bar g(\phi) = {2\over \ln k} 
    \ln \left(1+k-(1-k)\tilde g(\phi)\over 2\right) -1.
\ee
Using the latter relation, the equilibrium concentration profile
can also be rewritten as
\be
\label{eq:c0profile}
c_0(\phi)=c_l {1+k - (1-k)\tilde g(\phi)\over 2} = 
{c_s+c_l \over 2} + \tilde g(\phi) {c_s-c_l \over 2}. \label{ceqbis}
\ee
The physical meaning of the two interpolation functions is
hence completely transparent: $\bar g$ interpolates the internal
energy [Eq. (\ref{einterpolant})], and as a consequence the chemical 
potentials [Eqs. (\ref{mueq}) and (\ref{ceq})],
whereas $\tilde g$ interpolates the entropy density 
[Eq. (\ref{sinterpolant})] and, as
a consequence of Eq.~(\ref{gcond}), the concentration
[Eq. (\ref{ceqbis})].

If Eq.~(\ref{gcond}) is satisfied, the right-hand side of
Eq.~(\ref{phieq}) vanishes, and the solution for the equilibrium
profile of $\phi$ is the usual hyperbolic tangent,
$\phi_0(x)=-\tanh(x/\sqrt{2} W)$, where 
$W=(\sigma/H)^{1/2}$ measures the width of the diffuse interface.
Furthermore, the surface tension is defined as the excess of the 
grand potential $\omega=f - \mu c$, integrated through the interface, 
that is, $\gamma = \int_{dx} \omega(x) - \omega_E$. 
Because condition (\ref{gcond}) is equivalent to require
$\partial_\phi f_{AB}(\phi_0,c_0,T)=0$, under this condition
$f_{AB}(\phi_0,c_0,T)$ is independent of $x$ and equals
its bulk phase values $f_\nu(c_\nu,T)$. Since the latter enter
the expression for the equilibrium grand potential $\omega_E$ as
given by Eq. (\ref{ct2}), the contribution 
of $f_{AB}$ to $\omega(x)-\omega_E$ is zero. 
Thus, only the two other
interface terms in Eq. (\ref{fint0}) contribute. Taking into
account that both contribute the same amount (equipartition relation), 
we have $\omega(x) - \omega_E = H[1-\phi_0(x)^2]^2/2$, and hence
the surface tension is 
\be
\label{eq:surftension}
\gamma = IWH
\ee
with $I=2\sqrt{2}/3$. Like in the sharp-interface model of 
Sec. \ref{sharp}, $\gamma$ is independent of solute 
concentration and temperature. Let us stress again that
this property is only achieved if condition (\ref{gcond})
is satisfied. Otherwise, Eq.~(\ref{eq:surftension}) 
is replaced by a more complicated expression which 
contains the impurity concentration, and which needs in
general to be calculated numerically. A drawback of
this more complicated expression is that the dependence of $\gamma$
on concentration along the interface cannot be 
chosen independently of the value of $W$. 
This feature leads to an unphysically large variation
of $\gamma$ with concentration for computationally
tractable mesoscopic values of $W$. 
Eq. (\ref{eq:surftension}) yields a concentration-independent
expression for $\gamma$ that is free of this limitation.
Moreover, the fact that the
equilibrium profile remains a hyperbolic
tangent for arbitrary values of the concentration makes
the relationship between phase-field and sharp-interface
parameters obtained from the thin-interface analysis
independent of the value of the local 
concentration. This, in turn, avoids spurious kinetic corrections
that are present otherwise.

Once we have found a convenient relation between
$\tilde g(\phi)$ and $\bar g(\phi)$, we come back to the
complete dynamical model. 
The relations we have found in equilibrium can now be used to 
obtain two particularly simple
forms of the phase-field equation {\em out of equilibrium}.
For the first, we remark that Eq.~(\ref{gcond}) implies that
$\tilde g'(\phi_0) c_l (1-k) = -\bar g'(\phi_0) \ln k\, c_0$,
and therefore the function $\tilde g$ can be eliminated in
favor of the phase-dependent equilibrium concentration $c_0(\phi,T)$
and the function $\bar g$. Dividing Eq. (\ref{pa1}) by $H$,
we obtain
\be
\tau\frac{\partial \phi}{\partial t} = W^2\nabla^2\phi
+\phi-\phi^3+\frac{RT_m(T-T_m)}{2v_0Hm}\,\ln k\,\bar g'(\phi) 
    \left[c-c_0(\phi,T)\over c_l(T)\right],
\ee
with $\tau = 1/[K_\phi(T) H]$; the driving 
force is the local supersaturation. The temperature dependence
of $\tau$ will be addressed later in section IV.C.

The second possibility is to rewrite the phase-field equation
in terms of the dimensionless variable
\be
\label{udef}
u=\frac{v_0}{RT_m}(\mu-\mu_E)=\ln\left(c/c_l^0\right)
-\frac{\ln k}{2}\left(\bar g(\phi)+1\right) =
  \ln\left(2c\over c_l^0[1+k-(1-k)\tilde g(\phi)]\right),
\ee
which measures the departure of the chemical potential
from its value $\mu_E(T_0)$ for a flat interface 
at the equilibrium liquidus temperature $T_0$ (and
liquid concentration $c_l^0= c_l(T_0)$).
Then, it is preferable to eliminate $\bar g(\phi)$ in 
favor of $\tilde g(\phi)$. The result is the form
used in Ref.~\cite{KarmaPRL},
\be
\tau\frac{\partial \phi}{\partial t} = W^2\nabla^2\phi
+\phi-\phi^3-\frac{\tilde \lambda}{1-k} \tilde g'(\phi)
\left(e^u - 1 - \frac{T-T_0}{mc_l^0} \right) \label{prl1},
\ee
where we have defined the constant
\be
\tilde \lambda = \frac{RT_m(1-k)^2c_l^0}{2v_0H} =
                 \frac{L \Delta T_0}{2HT_m} ,
\ee
where we recall that $\Delta T_0=|m|(1-k)c_l^0$ is the freezing
range. Note that the parameter $H$ can be expressed in terms
of the surface tension, $H = \gamma/(IW)$. Then, we have
\be
\tilde\lambda = I \Delta T_0 W / (2 \Gamma), 
\label{lambdadef}
\ee
where 
$\Gamma$ is the Gibbs-Thomson constant of Eq.~(\ref{githoconst}).
Therefore, up to numerical constants, $\tilde\lambda$ is 
the dimensionless ratio of interface thickness times 
freezing range and the Gibbs-Thomson constant. It is
immediately clear that a variation of the interface
thickness corresponds to a change in $\tilde\lambda$.

\subsection{Non-variational formulations}
\label{nonvariational}

In spite of the theoretical appeal of a variational formulation,
relaxing the requirement that both Eqs. (\ref{pa1}) and (\ref{pa2})
derive from a single functional $F$ yields more flexibility.
In particular, this extra freedom can then be used to cancel out 
spurious effects.

\paragraph{Non-variational formulation without antitrapping current.}
\label{withoutanti}

In the last form proposed in the previous section, the 
interpolation function $\tilde g(\phi)$ enters the model not only
in the evolution equation for the phase field [Eq. (\ref{prl1})],
but also in that for the impurity, Eq. (\ref{pa2}), through the
change of variable Eq. (\ref{udef}). Whereas the condition
$\tilde g'(\pm 1)=0$ is necessary in the equation of motion for
$\phi$ to ensure that $\phi=\pm 1$ are the equilibrium solutions 
for arbitrary $u$ and $T$, no such condition is needed in the 
equation for the impurity. This suggests
replacing $\tilde g(\phi)$ in the definition of $u$ Eq. (\ref{udef})
by another function $h(\phi)$ which not necessarily satisfies
$h'(\pm 1)=0$, but still has the same limits $h(\pm 1)=\pm 1$:
\be
u = \ln\left(2c\over c_l^0[1+k-(1-k)h(\phi)]\right).
\ee
Thus, the equilibrium properties derived in the last section are
preserved; note, however, that the equilibrium {\em concentration
profile} $c_0(\phi)$ is modified because $h(\phi)$ replaces $\tilde g(\phi)$
in Eq.~(\ref{eq:c0profile}), yielding $c_0(\phi) = c_l^0[1+k-(1-k)h(\phi)]/2$. 
In practice, this allows the simple choice $h(\phi)=\phi$,
for which the equilibrium concentration profile has the lowest
possible gradients, and convergence of the simulations 
can hence be achieved for a coarser mesh \cite{KarRap}.

\paragraph{Non-variational formulation with antitrapping current.}
\label{withanti}

Albeit now $h(\phi)$ and $q(\phi)$ are completely free functions
which purely need to interpolate from 
+1 to -1 and from 0 to 1 respectively,
this does not yet provide enough freedom to cancel the
three spurious effects mentioned in the introduction. To
achieve this goal, we add an extra term in the model 
equations to specifically cancel one of them.
The extra interpolation function contained in this new term 
provides the necessary third degree of freedom to make all
three effects vanish.

We specifically target the solute trapping effect. This
occurs when solute atoms or molecules cannot escape
the advancing solidification front fast enough to maintain
local equilibrium at the interface. The characteristic interface 
velocities where solute trapping becomes important can be estimated
by comparing the time of advance by one interface thickness,
$W/V$, and the time it takes for the solute to diffuse
through the interface, $W^2/D$. The result is $V\sim D/W$,
and hence the critical speed depends on the interface
thickness. Since we ultimately want to simulate solidification
with diffuse interfaces that are orders of magnitude larger
than the real solid-liquid interfaces, solute trapping sets in
for much lower speeds than in reality. 

To eliminate this
interface-thickness effect, we introduce a supplementary
current in the equation for the solute concentration, the
{\em antitrapping current}. Its purpose is to transport
solute atoms from the solid to the liquid. Therefore, it
has to fulfill a number of properties. First, it must
be proportional to the speed of the interface, and hence
to $\partial_t\phi$. Next, it must be directed
from the solid to the liquid, that is, along the unit
normal vector $\hat n$, which in terms of the phase field can be 
expressed as (up to higher order corrections in the interface thickness)
$\hat n = - \vec\nabla \phi/|\vec\nabla\phi|$. Furthermore,
it must be proportional to the interface thickness $W$,
and to the local concentration difference between solid and liquid.
In contrast, we do not know {\em a priori} the profile of
the current function through the interface. The time derivative
of the phase field $\partial_t\phi$ is sensibly different from zero
only in the interface regions and induces a certain antitrapping
current profile. Additional freedom may be gained by allowing
for a shape function $a(\phi)$ that must be appropriately chosen
in order to obtain the correct thin-interface limit. 

In summary, we write
\be
j_{at} = a(\phi) W (1-k) c_l^0 e^u \frac{\partial\phi}{\partial t} \hat n = 
  - a(\phi) W (1-k) c_l^0 e^u 
    \frac{\partial\phi}{\partial t} {\vec \nabla \phi\over |\vec\nabla\phi|},
\ee
and the equation for the concentration becomes
\be
{\partial c\over \partial t} = 
    \vec{\nabla}\left(D\tilde q(\phi)c\vec{\nabla} u - j_{at}\right).
\ee
Note that the latter no longer derives from a functional $F$,
even if such a functional is allowed to be different from that
giving rise to the equation of motion for $\phi$. 

\paragraph{Formulation in terms of dimensionless supersaturation 
and relation with pure substance model.}
\label{uparamdiffuse}

It turns out to be advantageous for the subsequent asymptotic
analysis to make another change of variables in order to 
bring the equations in a form that is close to those analyzed 
in Refs.~\cite{KarRap,Alm}. To this end, we introduce the 
diffuse-interface extension $U(\phi)$ of the dimensionless 
supersaturation $U$ in Eq. (\ref{udefsharp}), now defined 
in the whole system,
\be
\label{udefdiffuse}
U = {e^{u}-1\over 1-k}.
\ee
Furthermore, we fix now the interpolation function $\tilde g$ to be
\be
\tilde g(\phi)=\frac{15}{8}\left(\phi-\frac{2\phi^3}{3}+\frac{\phi^5}{5}\right),
\ee
define new interpolation functions
\be
q(\phi) = \tilde q(\phi){1+k-(1-k)h(\phi)\over 2},
\ee
\be
g(\phi) = {8\over 15} \tilde g(\phi)=
   \left(\phi-\frac{2\phi^3}{3}+\frac{\phi^5}{5}\right)
\ee
and transform the equation for $c$ into one for $U$. 
Taking into account that $T(z)=T_0 + G(z-V_pt)$ and the
temperature-dependent relaxation time 
$\tau=\tau_0 [1 - (1-k)(z-V_pt)/l_T]$ discussed later
in section IV.C, the final set of equations is
\begin{eqnarray}
\tau_0 \left[1 - (1-k)\frac{z-V_pt}{l_T}\right]
\frac{\partial \phi}{\partial t} &=&W^2\nabla^2\phi
+\phi-\phi^3-\lambda\, g'(\phi) \left(U+ \frac{z-V_pt}{l_T}\right) 
\label{fin1},\\
\left(\frac{1+k}{2} - \frac{1-k}{2}h(\phi)\right) 
   \frac{\partial U}{\partial t}&=& 
\vec \nabla \cdot \left( D q(\phi)
\,\vec\nabla U + a(\phi) W \left[1+(1-k)U\right] 
\frac{\partial \phi}{\partial t}\,\frac{\vec \nabla \phi}
{|\vec \nabla \phi|}\right)\nonumber\\
& &\mbox{} + \left[1+(1-k)U\right] 
   \frac{1}{2}\frac{\partial h(\phi)}{\partial t}
\label{fin2},
\end{eqnarray}
where
\be
\lambda =\frac{15}{8} \tilde\lambda .
\label{lambda}
\ee
With these choices, the phase field equation [Eq. (\ref{fin1})] 
becomes identical to the one analyzed in Ref. \cite{KarRap}.
One important advantage of this formulation is that the
special case of a constant concentration jump can be
recovered without any difficulty by setting $k=1$, whereas
in the formulation with the variable $u$, the limit $k\to 1$
has to be treated with some care. Hence, the model
of Ref. \cite{Alm} is contained as a special
case of Eqs.~(\ref{fin1}) and (\ref{fin2}), for $k=1$.

\section{Thin-interface analysis}
\label{asym}

\subsection{Introductory remarks}

The goal of the matched asymptotic analysis is to relate the
phase-field model [Eqs. (\ref{fin1}) and (\ref{fin2})]
to a free-boundary problem. In particular, we would like to
recover that of Eqs. (\ref{fbU1}--\ref{fbU3}). The principle
is to choose the interface width much smaller than any physically
relevant length scale. This difference in scale can be exploited
for a perturbation expansion, in which the solution on the 
{\em outer scale} of the transport field is first assumed to
be known. For a given point of the interface, this fixes the
local velocity and curvature. The reaction of the diffuse
interface to this ``forcing'' can then be calculated on the
{\em inner scale} of the interface width, which yields a boundary
condition for the diffusion field on the outer scale. The
matching of both solutions then provides the link between
``outer'' (physical) and ``inner'' (phase-field) parameters.

Two different perturbation schemes have been used. The ``classic''
one, developed by Langer, Caginalp, and others, uses the ratio
of interface thickness and capillary length, $\epsilon=W/d_0$,
as an expansion parameter. Later, Karma and Rappel remarked that
the physically relevant length scales for the outer problem are
not the capillary length, but rather the diffusion length $D/V$
or a local radius of curvature $\rho$. Calculations performed
with the expansion parameter $p=WV/D$ for the symmetric model
of solidification ($D_s=D_l$, or $q(\phi)=1$) yield, to first order
in $p$, a new expression for the interface kinetic coefficient
that contains a finite-interface thickness correction. This
has allowed a tremendous gain in calculation power, since
much larger $W$, including $\epsilon\gg1$, can be used.
It was also shown that this correction can be obtained in
a second-order expansion in $\epsilon$ \cite{KarRap,Alm}.

Here, we will follow the classic scheme and present the asymptotic 
analysis for our model up to second order in $\epsilon$. While
$\epsilon$ is not necessarily small, this method yields all 
important correction terms at second order, while other 
schemes need to include some third order terms. The reasons
for this, as well as the conditions of convergence of the
expansion in $\epsilon$, can be better appreciated in the 
light of the formal results given below, and a discussion 
of these points is therefore deferred to Sec. IV.D.

To perform the analysis, it is advisable to use a dimensionless
version of the equations. We will use as unit length the capillary 
length $d_0$ and as unit time $d_0^2/D$. Without loss of generality,
we set $t=0$ (which amounts to a shift of reference frame) such
that the term $V_pt$ drops out. Furthermore, we remark that 
from the definitions of Eqs.~(\ref{d0def}), (\ref{lambdadef}), 
and (\ref{lambda}), we obtain
\be
d_0 = a_1 \frac{W}{\lambda}\label{eqd0}
\ee
with $a_1 = I/J$, where $J=g(+1)-g(-1)$. For our choices of
functions, $I=2\sqrt{2}/3$ and $J=16/15$, such that $a_1=5\sqrt{2}/8$.
Therefore, $\lambda$ can be eliminated of the equations in 
favor of $a_1\epsilon$. The result reads
\begin{eqnarray}
\alpha\epsilon^2 \partial_t \phi &=&\epsilon^2\nabla^2\phi
- f'(\phi)-a_1\epsilon g'(\phi) \left(U+\nu z\right) 
\label{dimless1},\\
\left(\frac{1+k}{2} - \frac{1-k}{2}h(\phi)\right) 
   \partial_t U &=& 
\vec \nabla \cdot \left( q(\phi)
\,\vec\nabla U + \epsilon a(\phi) \left[1+(1-k)U\right] 
\frac{\partial \phi}{\partial t}\,\frac{\vec \nabla \phi}
{|\vec \nabla \phi|}\right)\nonumber\\
& &\mbox{} + \left[1+(1-k)U\right] 
   \frac{\partial_t h(\phi)}{2}
\label{dimless2},
\end{eqnarray}
where we have introduced the dimensionless parameters 
$\nu\equiv d_0/l_T$ and $\alpha\equiv D\tau/W^2$,
and defined the double-well function $f=-\phi^2/2+\phi^4/4$. 
We will assume that $\epsilon$ is the only small
parameter and consider all other parameters of $O(1)$. Note that 
$\nu=d_0/l_T$ is a physical parameter that is typically small, but 
independent of the computational parameter $\epsilon$, and therefore,
$\nu=O(1)$. The parameter $\alpha$ depends on the choice of
$\tau$; we consider it to be of $O(1)$ in order to avoid
neglecting any important terms. Our conclusions remain
valid if $\alpha$ is of order $\epsilon$ or smaller.

For comparison, we also adimensionalize
the free-boundary problem we would like to recover,  
Eqs.~(\ref{fbU1}--\ref{fbU3}),
using the above rescaling of space and time:
\begin{eqnarray}
\partial_t U &=&\nabla^2 U\quad{\rm (liquid)}\;,\label{fbnodim1} \\
\left[1+(1-k)U \right] v_n&=&-\partial_nU|^+
  \quad{\rm (interface)}\;,\label{fbnodim2}\\
U &=&-\kappa-\tilde\beta v_n-\nu z \quad{\rm (interface)}\;,\label{fbnodim3}
\end{eqnarray}
where $\kappa=d_0{\cal K}$ and $v_n=d_0V_n/D$ are the dimensionless 
interface curvature and normal velocity,
and $\tilde\beta = \beta D/d_0 $, the dimensionless kinetic
coefficient. In the following, we will show how to recover this
model as closely as possible by choosing specific forms for the
functions $q(\phi)$, $h(\phi)$ and $a(\phi)$.

\subsection{Matched asymptotic expansions}
We make a perturbation analysis in powers of $\epsilon$ in
the inner region,
\begin{eqnarray}
\label{formalexpansion1}
\phi & = & \phi_0 + \epsilon \phi_1 + \epsilon^2 \phi_2 + \ldots \, , \\
\label{formalexpansion2}
U & = & U_0 + \epsilon U_1 + \epsilon^2 U_2 + \ldots \, ,
\end{eqnarray}
and similarly in the outer region, 
$\tilde \phi = \tilde \phi_0 + \epsilon \tilde\phi_1 + \ldots$,
$\tilde U = \tilde U_0 + \epsilon \tilde U_1 + \ldots$. In the
outer region, Eqs.~(\ref{dimless1}) and (\ref{dimless2}) can
be expanded in powers of $\epsilon$ in a straightforward manner.
Since we have $g'(\pm 1) = 0$, $\tilde\phi=\pm 1$ are stable solutions
for the phase-field equations to all orders in $\epsilon$ for 
any value of $U$ and $z$. Therefore, the outer solution for the 
phase field is simply a step function, and the field $\tilde U$ 
obeys the diffusion equation to all orders,
\be
\partial_t \tilde{U}= q(\pm 1) \nabla^2 \tilde{U},
\ee
where we recall that $q(1)=0$ and $q(-1)=1$ for the one-sided model.
Also, note that 
the local equilibrium condition for the concentrations will be satisfied
at all orders to which $\tilde U$ is continuous across the interface.

In the dimensionless equations, the Laplacian of the phase field 
comes with a prefactor $\epsilon^2$, which leads to the two distinct
constant $\phi$ solutions in the outer region on the two sides
of the interfaces. In the inner region, the phase field varies
smoothly. Equation (\ref{dimless1}) tells one that, for $\epsilon \to 0$, 
this is only possible if such a variation takes
place precisely on a scale of $O(\epsilon)$, which renders 
$\nabla^2 \phi = O(\epsilon^{-2})$ and 
invalidates the counting of orders used above.
To compute the inner solution, we therefore must rescale
the coordinate normal to the interface. We introduce
the curvilinear coordinates in the reference frame of the 
interface $r$ (signed distance to the level line $\phi=0$) and 
$s$ (arclength along the interface), and
define the rescaled coordinate $\eta\equiv r/\epsilon$. 
Standard formulas of differential geometry yield [see e.g. \cite{Folchetal}]
\begin{eqnarray}
\nonumber
\partial_t &=& -\epsilon^{-1}v_n\partial_\eta + d_t - v_t\partial_s 
             + O(\epsilon), \\
\nonumber
\nabla^2 &=& \epsilon^{-2}\partial^2_\eta + \epsilon^{-1}\kappa\partial_\eta
           -\kappa^2\eta\partial_\eta + \partial^2_s + O(\epsilon), \\
\nonumber
\vec\nabla\cdot(q\vec\nabla) &=& 
\epsilon^{-2}\partial_\eta(q\partial_\eta)
+\epsilon^{-1}\kappa q\partial_\eta -\kappa^2 q\eta\partial_\eta
+\partial_s (q\partial_s ) + O(\epsilon), \\
\nonumber
z &=& z_i + \epsilon (\hat n\cdot\hat z) \eta, \\
\nonumber
- \frac{\vec\nabla\phi}{|\vec\nabla\phi|} &=& 
   \hat n\left[1+O(\epsilon^2)\right] + \hat s O(\epsilon), \\
\nonumber
\vec\nabla \cdot \vec a &=& \epsilon^{-1}\partial_\eta(\hat n\cdot\vec a) 
+\partial_s(\hat s\cdot\vec a) +\kappa\hat n\cdot\vec a + O(\epsilon),  
\end{eqnarray}
where $\vec a$ is a vector function of the fields, $v_n (v_t)$ are the
dimensionless normal (tangential) velocity of the interface, $z_i$ its
dimensionless $z$ position, and $d_t$ is the time derivative 
at fixed $r$ and $s$. 

Since changes in the arclength $s$ amount to a 
re-parametrization, we neglect terms in $v_t$ without loss of generality.
We will also neglect the operators $d_t$. 
This amounts to the assumption that the interface 
follows adiabatically the changes in the forcing.
For the phase field $\phi$, this approximation is always justified, since 
this field has an approximately stationary kink shape moving with 
the interface (this will be explicitly checked by computing $\phi$ at
lowest order in $\epsilon$,  which turns out to be a function of $\eta$
only). For the diffusion field $U$, it can be seen from Eq.~(\ref{fbnodim3})
that $d_t U \neq 0$ originates from variations with time of the
interface curvature, velocity, and position. The variations of the
latter occur generally on the slow time scale of solute redistribution
transients, $D/V_p^2$, and are therefore negligibly small. The 
characteristic time scale for variations of the curvature and velocity
is $R/V_n$, where $R=1/{\cal K}$ is the local radius of curvature, since 
this is the time the interface needs to move by once its radius of curvature. 
Therefore, the curvature and velocity terms in $d_t U$ are of order 
$v_n \kappa (\kappa + \tilde \beta v_n)$. Since $\kappa$ and $v_n$
themselves are small quantities, $d_t U$ is much smaller than other
terms of order $v_n \kappa$ which will appear in the calculation below,
and can hence safely be dropped.

We substitute the above expressions into Eqs.~(\ref{dimless1}) 
and (\ref{dimless2}) to obtain
\begin{eqnarray}
\nonumber
&\;\;&\;\partial^2_\eta \phi - f'(\phi) \\ 
\nonumber
&+&\epsilon\left[(\alpha v_n+\kappa)\partial_\eta \phi - a_1 g'(\phi)
\left(U + \nu z_i \right)\right] \\
\label{eq:pfin}
&+&\epsilon^2\left[\partial^2_s \phi - \kappa^2\eta\partial_\eta \phi
-  a_1\nu (\hat n\cdot\hat z) \eta g'(\phi)\right]
= {\cal O}(\epsilon^3),
\end{eqnarray} 
\begin{eqnarray}
\nonumber
&\;\;&\;\epsilon^{-2}\partial_\eta(q\partial_\eta U)\\
\nonumber
&+&\epsilon^{-1}\left\{\left[ v_n 
  \left(\frac{1+k}{2} - \frac{1-k}{2}h(\phi)\right)+\kappa q\right] 
   \partial_\eta U \right. \\
\nonumber
& & \qquad\qquad\mbox{}
+ v_n \partial_\eta\left\{ a\left[1+(1-k)U\right] 
   \partial_\eta\phi \right\}
   - \frac{v_n}{2}\left[1+(1-k)U\right]\partial_\eta h\Biggr\} \\
\label{eq:diffusionin}
&+& \epsilon^0 \left\{ \partial_s(q\partial_s U)-\kappa^2\eta q\partial_\eta U
+a v_n\kappa \left[1+(1-k)U\right] \partial_\eta\phi\right\}
= {\cal O}(\epsilon)
\end{eqnarray}
and solve them order by order in $\epsilon$. The matching to
the outer expansion is trivial for $\phi$ since the outer solution
is just a step function. For $U$, the matching conditions read
\begin{eqnarray}
\nonumber
&&
\lim_{\eta=\pm\infty}\left[ U_0(\eta,s) - \tilde{U}_0|^\pm(s)\right] = 0, \\
\nonumber
&&
\lim_{\eta=\pm\infty} \left[U_1(\eta ,s) -
     \left(\tilde{U}_1|^\pm(s)+\eta\partial_r\tilde{U}_0|^\pm(s)\right)\right] = 
0,\\
\label{eq:matching}
&&
\lim_{\eta=\pm\infty} \left[U_2(\eta ,s) -
     \left(\tilde{U}_2|^\pm(s)+\eta\partial_r\tilde{U}_1|^\pm(s)+
      (\eta ^2/2)\partial^2_r\tilde{U}_0|^\pm(s)\right)\right] = 0,
\end{eqnarray}
where $|^\pm$ means that the outer field and its derivatives
are evaluated at the interface, coming from either the $+$ 
(liquid) or the $-$ (solid) side. As a consequence,
\begin{eqnarray}
\nonumber
\lim_{\eta=\pm\infty}\partial_\eta U_0(\eta,s)&=&
\lim_{\eta=\pm\infty}\partial^2_\eta U_1(\eta,s)=0, \\
\nonumber
\partial_r\tilde{U}_0|^\pm(s) &=&
\lim_{\eta=\pm\infty}\partial_\eta U_1(\eta,s),
\\
\label{eq:termmatch}
\partial_r\tilde{U}_1|^\pm(s) &=&
\lim_{\eta=\pm\infty} \left[ \partial_\eta U_2(\eta,s,t)
   -\eta \partial^2_r\tilde{U}_0|^\pm(s) \right].
\end{eqnarray}
This matching will provide the boundary conditions on the interface 
for the outer concentration. We now proceed to solve the inner 
equations order by order.

\paragraph*{Gibbs--Thomson relation.}
\label{gt}
Equation (\ref{eq:pfin}) at order $\epsilon^0$,
\begin{equation}
\label{eq:pfin0}
\partial^2_\eta\phi_0-f'(\phi_0)=0
\end{equation}
yields, with the boundary conditions $\phi_0 \to -1$ for $\eta\to +\infty$ 
and $\phi_0 \to 1$ for $\eta\to -\infty$ set by the matching to the outer
solution, the zeroth order solution
\be
\phi_0(\eta) = - {\rm tanh} \frac{\eta}{\sqrt{2}} \, .
\ee 
In turn, Eq. (\ref{eq:diffusionin}) at order $\epsilon^{-2}$ becomes
\be
\partial_\eta\left(q(\phi_0)\partial_\eta U_0\right) = 0 \, ,
\ee
which can be integrated once to yield
$q(\phi_0) \partial_\eta U_0=A_0(s)$.
Taking the $\eta \rightarrow \pm \infty$ limit according to
Eq. (\ref{eq:termmatch}) we find $A_0(s)=0$, and therefore
\begin{equation}
\label{eq:uin0}
U_0=\bar{U}_0(s).
\end{equation}
To fix this constant, in turn, we consider Eq.~(\ref{eq:pfin}) at order 
$\epsilon$,
\begin{equation}
\label{eq:pfin1}
{\cal L}\phi_1 = a_1 g'(\phi_0) \left(U_0 + \nu z_i \right)
-(\alpha v_n+\kappa)\partial_\eta \phi_0,
\end{equation}
where ${\cal L}\equiv \partial^2_\eta-f''(\phi_0)$ is a linear
differential operator. Since the partial derivative with respect 
to $\eta$ of Eq. (\ref{eq:pfin0}) is ${\cal L}\partial_\eta \phi_0 = 0$,
$\partial_\eta\phi_0$ is an eigenfunction of $\cal L$ with 
eigenvalue zero. Therefore, the solvability condition for 
the existence of a nontrivial solution $\phi_1$ reads
\begin{equation}
\label{eq:solvability1}
  a_1 \left({\bar U}_0 + \nu z_i \right) J + (\alpha v_n+\kappa) I = 0,
\end{equation}
where $J\equiv \int_{+\infty}^{-\infty} g'(\phi_0) \partial_\eta \phi_0 d\eta
= g(+1)-g(-1)$ and $I\equiv\int_{-\infty}^{+\infty}
\left(\partial_\eta \phi_0\right)^2 d\eta$.
Since $I$ and $J$ are the same constants that have been used
to define $a_1 = I/J$, we obtain
\be
{\bar U}_0 = -\nu z_i - \alpha v_n - \kappa,
\label{a1obtain}
\ee
which is identical to the Gibbs-Thomson condition of the free 
boundary problem, Eq.~(\ref{fbnodim3}), with 
$\tilde\beta\equiv\tilde\beta_0=\alpha$.

This is the ``classic'' result for the kinetic coefficient
in the sharp-interface limit. To obtain the thin-interface
correction, we repeat the same procedure at next order.
Thanks to Eq. ~(\ref{eq:uin0}) we can drop the terms in 
$\partial_\eta U_0$ arising in
Eq.~(\ref{eq:diffusionin}) at order $\epsilon^{-1}$, to obtain
\begin{equation}
\label{eq:diffusionin1}
\partial_\eta\left[q(\phi_0)\partial_\eta U_1\right]=
-v_n 
\partial_\eta\left\{a(\phi_0)\left[1+(1-k)U_0\right]\partial_\eta\phi_0\right\}
+ \frac{v_n}{2}\left[1+(1-k)U_0\right] \partial_\eta h(\phi_0),
\end{equation}
and integrate it once with respect to $\eta$ to yield
\begin{equation}
\label{eq:intdiffusionin1}
q(\phi_0)\partial_\eta U_1 = v_n \left[1+(1-k)U_0\right]
\left[h(\phi_0)/2 - a(\phi_0)\partial_\eta\phi_0\right] + A_1(s),
\end{equation}
where $A_1(s)$ is an integration constant. The latter can be fixed by
considering the limit $\eta\to-\infty$. In fact, the left-hand
side represents the diffusion current, which vanishes inside
the bulk solid. Since the antitrapping current must also vanish 
and $h(1)=1$, we find $A_1(s)=- (v_n/2) \left[1+(1-k)U_0\right]$. 
Substituting it back into Eq. (\ref{eq:intdiffusionin1} and 
integrating the latter once more between 0 and $\eta$, we find
\begin{equation}
\label{eq:uin1}
U_1 = {\bar U}_1 + 
  \frac{v_n}{2} \left[1+(1-k)U_0\right] \int_0^\eta p[\phi_0(\xi)] \,d\xi,
\end{equation}
where ${\bar U}_1$ is the value of $U_1$ at the interface ($\eta=0$), and
\be
p(\phi_0) = \frac{h(\phi_0)-1 - 2a(\phi_0)\partial_\eta\phi_0}{q(\phi_0)}.
\label{pdef}
\ee
The profile $U_1$ therefore depends on the choice of the functions
$q(\phi)$, $h(\phi)$ and $a(\phi)$. Note that both the denominator
and the numerator tend to zero when $\eta\to -\infty$. It is important
here to remark that we need to require $p\to 0$ in this limit, since
otherwise $U_1$ diverges, which makes a matching to the outer solution
($U$ is constant in the solid) impossible. In fact, this property
makes the standard asymptotic expansion inconsistent. A careful
analysis, carried out in the appendix, shows that in this case
a term of order $p\log p$ (with $p=WV/D$ the interface Peclet number)
appears in the interface kinetics, which makes the convergence of the 
model to the sharp-interface limit very slow. This term appears, 
for example, in the standard formulation of the one-sided model that 
has been widely used \cite{Wheeler,Warren95}. In order to avoid this 
phenomenon, we will require in the following $p(\phi)\to 0$ for $\phi\to 1$,
that is, the numerator must vanish more rapidly than the denominator.

Under this condition, we may fix the constant $\bar{U}_1$ by 
considering Eq.~(\ref{eq:pfin}) at order $\epsilon^2$,
\begin{eqnarray}
\nonumber
{\cal L}\phi_2 & = & \frac{f'''(\phi_0)}{2} \phi_1^2 
   - (\alpha v_n +\kappa) \partial_\eta \phi_1
   + a_1 g'(\phi_0) U_1
   + g''(\phi_0) \phi_1 a_1 (U_0 + \nu z_i) \\ 
   & & \; \mbox{}   + \kappa^2\eta\partial_\eta \phi_0
     + g'(\phi_0) a_1 \nu (\hat n\cdot \hat z) \eta
\label{eq:pfin2}
\end{eqnarray}
where we have used $\partial_s \phi_0=0$. In this expression
appears the first order correction to the phase-field, $\phi_1$,
which is the solution of the differential equation obtained by
substitution of Eq. (\ref{eq:solvability1}) into Eq. (\ref{eq:pfin1}),
\begin{equation}
\label{eq:diffeqphi1}
{\cal L}\phi_1 = -(\alpha v_n+\kappa)
\left(a_1 g'(\phi_0) + \partial_\eta \phi_0\right),
\end{equation}
with the boundary conditions $\phi_1(\eta \rightarrow \pm\infty)=0$
imposed by the matching to the outer solution.
Clearly, $\phi_1$ equals $\alpha v_n+\kappa$ times a function only of $\eta$, 
so that, when substituted into Eq. (\ref{eq:pfin2}), 
it would yield $(\alpha v_n+\kappa)^2$ contributions to $\bar U_1$.
There are essentially two ways to avoid this problem. The first
would be to choose $g$ such that
\begin{equation}
\label{asg}
g'(\phi_0) = - \partial_\eta \phi_0/a_1,
\end{equation}
which makes $\phi_1$ vanish. For our standard quartic double-well
potential which yields $\partial_\eta \phi_0 = (1-\phi_0^2)/\sqrt{2}$,
the corresponding $g$ function is a third-order polynomial that
has been widely used. However, we have chosen here a different
function, and many calculations have also been performed with 
yet other interpolation functions, such that this condition
is too restrictive. The second way out is to use the symmetry
properties of the involved functions. For any symmetric double-well
function (that is, $f(-\phi)=f(\phi)$), the equilibrium profile
is odd in $\eta$, $\phi_0(-\eta)=-\phi_0(\eta)$,
and its derivative is even. If $g$ is chosen to be odd
in $\phi$, $g(-\phi)=-g(\phi)$, then $g'(\phi_0)$ is also even 
in $\eta$. Therefore, the entire right-hand side of 
Eq.~(\ref{eq:diffeqphi1}) is even. Since $\cal L$ is also an even 
operator, $\phi_1$ must be even, and its derivative $\partial_\eta\phi_1$ odd.
Given that the solvability condition is obtained by multiplying the
right-hand side of Eq.~(\ref{eq:pfin2}) by $\partial_\eta \phi_0$,
an even function, and integrating from $-\infty$ to $+\infty$,
the contribution of all odd terms vanishes. The only remaining
is the one that contains $U_1$, and the solvability condition reads
\begin{equation}
\label{eq:solvability2}
\frac{v_n}{2} \left[1+(1-k)U_0\right] K - J \bar U_1 = 0
\end{equation}
where we have expressed $U_1$ according to Eq.~(\ref{eq:uin1}), and
\begin{equation}
K = \int_{-\infty}^{+\infty} \,d\eta\,\partial_\eta \phi_0 g'(\phi^0)
    \int_0^\eta p(\phi_0(\xi)) \,d\xi .
\end{equation}
To obtain the desired result, namely an expression for $\tilde U_1$,
let us first remark that in the limit $\eta\to\infty$, Eq.~(\ref{eq:uin1})
yields $\partial_\eta U_1 = - v_n \left[1 + (1-k)U_0\right]$, which
is just the Stefan condition at lowest order. Using the matching
conditions $\lim_{\eta\to \pm\infty} \partial_\eta U_1 = \partial_r
\left.\tilde U_0\right|^\pm$, and
$\tilde{U}_1|^\pm = \lim_{\eta \rightarrow \pm \infty}
U_1(\eta) - \eta \partial_r \tilde{U}_0|^\pm$, we obtain
\begin{equation}
\tilde{U}_1|^\pm = - v_n \tilde\beta_1^\pm,
\end{equation}
\begin{equation}
\tilde\beta_1^\pm = - \left[1 + (1-k)U_0\right] \frac{JF^\pm + K}{2J}, 
\end{equation}
\begin{equation}
F^\pm \equiv \int_0^{\pm\infty} [p(\phi_0) - p(\phi^\pm)] d\eta.
\end{equation}
Note that $U$ will be continuous across the interface up to $O(\epsilon)$
if and only if $F^+ = F^- \equiv F$ (and hence 
$\tilde\beta_1^+=\tilde\beta_1^-$). 
Since $\tilde U = \tilde U_0 + \epsilon \tilde U_1$, the total kinetic
coefficient is 
\begin{equation}
\tilde \beta^\pm = \tilde\beta_0 +\epsilon\tilde\beta_1^\pm = 
   \alpha - \epsilon \left[1 + (1-k)U_0\right]\frac{K+JF^\pm}{2J}.
\label{eq:beta}
\end{equation}
The implications of this finding will be discussed below.

\paragraph*{Mass conservation.}
As already mentioned before, Eq.~(\ref{eq:uin1}) together
with the matching condition (\ref{eq:termmatch}) yields 
$\partial_r \tilde U_0|^- = 0$ and $\partial_r \tilde U_0|^+ = 
- v_n \left[1 + (1-k)\tilde U_0\right]$, which
is just the Stefan condition at lowest order.
In order to evaluate eventual corrections,
we proceed by calculating the normal gradients
at order $\epsilon$ using the matching condition
for $\partial_r \tilde{U}_1|^\pm$ in Eqs.~(\ref{eq:termmatch}).
The quantity $\partial^2_r \tilde{U}_0|^\pm$ can be evaluated
by remarking that the outer problem satisfies a simple diffusion
equation in a moving curvilinear coordinate system, and therefore
$[\partial_{rr}+(v_n+\kappa)\partial_r+\partial_{ss}]\tilde U_0 = 0$,
such that $\partial^2_r \tilde U_0|^\pm = 
-[(v_n+\kappa)\partial_r+\partial_{ss}]\tilde U_0|^\pm$.
To obtain $\partial_\eta U_2(\eta)$, 
Eq.~(\ref{eq:diffusionin}) is evaluated at $O(\epsilon^0)$
and integrated once from $0$ to $\eta$,
\begin{eqnarray}
& & q(\phi_0)\partial_\eta U_2 + q'(\phi_0)\phi_1\partial_\eta U_1
\nonumber \\
& + & \kappa\int_0^\eta\,d\xi\,q(\phi_0)\partial_\xi U_1 +
   v_n \int_0^\eta\,d\xi \left(\frac{1+k}{2} - \frac{1-k}{2}h(\phi_0)\right)
    \partial_\xi U_1
\nonumber \\
& + & v_n \left\{a'(\phi_0)\phi_1\left[1+(1-k)U_0\right]
              +a(\phi_0)(1-k)U_1\right\}\partial_\eta \phi_0
\nonumber \\
& + & v_n a(\phi_0)\left[1+(1-k)U_0\right] \partial_\eta \phi_1
      + v_n\kappa \left[1+(1-k)U_0\right] 
      \int_0^\eta \,d\xi\,a(\phi_0) \partial_\xi \phi_0
\nonumber \\
& - & \frac{v_n}{2} (1-k) \int_0^\eta \,d\xi\, U_1 \partial_\xi h(\phi_0)
       - \frac{v_n}{2}\left[1+(1-k)U_0\right] h'(\phi_0)\phi_1
\nonumber \\
& + & \partial_{ss} \tilde U_0 \int_0^\eta \,d\xi\, q(\phi_0) = A_2(s),
\label{horrible}
\end{eqnarray}
where $A_2(s)$ is an integration constant and we have taken into account 
that $\partial_\eta U_0 = \partial_s \phi_0 = 0$. Fortunately, we can
drop many terms of this long equation because we are only interested
in the limits $\eta\to\pm\infty$. In this limit, $\phi_1$ and 
$\partial_\eta \phi_0$ are exponentially small, such that all
terms containing them can be dropped, except when they appear
under an integral. The third term on the left-hand side of 
Eq.~(\ref{horrible}) can be rewritten using Eqs.~(\ref{eq:uin1}) 
and (\ref{pdef}) as
\begin{equation}
(3) = \frac{\kappa v_n}{2} \left[1+(1-k)U_0\right] 
       \int_0^\eta \,d\xi\, 
     \left[h(\phi_0) -1 -2a(\phi_0) \partial_\xi \phi_0\right],
\end{equation}
and it can be seen that the part proportional to $a(\phi_0)$ 
cancels out with the seventh term on the left-hand side. The 
remaining piece can be rewritten, using the Stefan condition
to lowest order, as
\begin{equation}
\lim_{\eta\rightarrow \pm\infty} [
(3) + (7)] = \kappa\eta \partial_r\tilde U_0|^\pm
      +  \frac{\kappa v_n}{2} \left[1+(1-k)U_0\right] 
       \int_0^\eta \,d\xi\, \left[h(\phi_0) - h(\mp 1)\right].
\end{equation}
Next, the remaining terms that contain $h$ can be grouped
and integrated to yield
\begin{equation}
(4) + (8) = v_n \left(\frac{1+k}{2} - \frac{1-k}{2}h(\phi_0)\right) U_1(\eta).
\end{equation}
Using the matching condition for $U_1$, 
$\lim_{\eta\to\pm\infty}U_1(\eta)=
\tilde U_1|^\pm + \eta \partial_r \tilde U_0|^\pm$, and the 
fact that $\lim_{\eta\to -\infty} q(\phi_0)\partial_\eta U_2 =0$,
we can obtain the constant $A_2$ from the limit $\eta\to -\infty$
of the entire Eq.~(\ref{horrible}),
\be
A_2 = \frac{\kappa v_n}{2} \left[1+(1-k)U_0\right]
\int_0^{-\infty} \,d\eta\, [h(\phi_0) - 1] +
v_n k \tilde U_1^- + \partial_{ss}\tilde U_0 
\int_0^{-\infty} \,d\eta\, q(\phi_0)  \; .
\ee
Next, $\lim_{\eta\to \infty} q(\phi_0)\partial_\eta U_2$ is
evaluated using the above result for $A_2$. Finally, with the 
help of the matching condition and the expression for 
$\partial^2_r \tilde U_0|^+$, we obtain
\begin{eqnarray}
\partial_r \tilde U_1|^+ & = & -\frac{\kappa v_n}{2} \left[1+(1-k)U_0\right] 
(H^+-H^-)
            - \partial_{ss}\tilde U_0 (Q^+-Q^-)  \nonumber\\
    & & \mbox{}
            - v_n (1-k) \tilde U_1^+ - v_n k (\tilde U_1^+ - \tilde U_1^-)
\label{eq:dru1}
\end{eqnarray}
with
\begin{equation}
H^\pm = \int_0 ^{\pm\infty} \,d\eta\,[h(\phi_0(\eta))-h(\phi^\pm)],
\label{eq:hdef}
\end{equation}
\begin{equation}
Q^\pm = \int_0 ^{\pm\infty} \, d\eta\, [q(\phi_0(\eta))-q(\phi^\pm)].
\label{eq:qdef}
\end{equation}
The first two terms on the right-hand side of Eq.~(\ref{eq:dru1}) are 
the announced finite interface thickness effects associated 
with interface stretching and surface diffusion; the third
is the expected first-order term that appears on the left-hand
side of the Stefan condition, Eq.~(\ref{fbnodim2}); finally, the
last one is a correction associated with a jump of $U$ through 
the interface. In total, the mass conservation condition for
the outer fields up to first order reads (recall that 
$\tilde U_0 = U_0$)
\begin{eqnarray}
\left[1+(1-k)(\tilde U_0+ \epsilon \tilde U_1)\right] v_n 
& = & -\partial_r (\tilde U_0 + \epsilon \tilde U_1) + \epsilon
  \biggl\{\frac{\kappa v_n}{2} \left[1+(1-k)\tilde U_0\right] (H^+-H^-) 
\nonumber\\
  & & \qquad \mbox{} + \partial_{ss}\tilde U_0 (Q^+-Q^-) \nonumber \\
  & & \qquad \mbox{} 
         + \frac{v_n^2 k}{2} \left[1+(1-k)\tilde U_0\right] (F^+-F^-)\biggr\}.
\label{eq:stefaneps}
\end{eqnarray}

\subsection{Discussion}

\paragraph*{Physical interpretation of the corrections:}
There are three corrections in $\epsilon$ to the 
classic free-boundary problem. The term proportional to 
$Q^+-Q^-$ describes the response of the interface to lateral 
concentration gradients, caused by variations of the curvature 
or the growth speed along the interface. For a diffuse 
interface, the resulting mass flow is smaller than in the 
bulk liquid on the liquid side, but larger than in the bulk 
solid on the solid side. If the two effects do not exactly 
compensate, a surface diffusion term needs to be 
included in the Stefan condition. The condition to make this
correction vanish is $Q^+=Q^-$, which can be shown to be 
exactly the same as Eq.~(\ref{SD}) in the introduction by 
taking into account that $q(\phi)=\tilde{q}(\phi) c_0 (\phi)/c_l$. 

Next, the term proportional to $H^+-H^-$ arises from the 
source term in the $U$ equation. If a positively curved 
interface moves forward, the liquid side of the interface 
is slightly longer than the solid side. Therefore, the 
source term on the liquid side is active over a larger 
area than the one on the solid side, and the integral of 
the source strength multiplied by the area over
which it is active is precisely given by the difference
$H^+-H^-$. If this quantity is non-vanishing, the interface
acquires a ``net impurity content'', that is, a source term
appears in the mass conservation condition when the length
of the interface changes, which is precisely the case if
the product $v_n\kappa$ is non-zero. This is the 
interface stretching correction, which vanishes
when $H^+=H^-$. In terms of the concentration, this
condition is identical to Eq.~(\ref{STR}).

Finally, the last correction involves a macroscopic
discontinuity in $U$ that is proportional to the velocity
$v_n$ and to $F^+-F^-$, and that appears in the boundary
conditions at the interface and in the Stefan condition, 
Eq.~(\ref{eq:stefaneps}). This is the solute trapping
term: since the concentrations on both sides of the
interface vary with velocity, they do not satisfy the
partition relation $c_s = k c_l$ out of equilibrium, or,
in other words, the solute rejection is velocity-dependent.
Since $U$ can be assimilated to a chemical potential, its
jump can be interpreted as resulting from a finite
interface mobility that leads to interface dissipation.
Note that both analogies are limited: whereas a ``physical''
dissipation is necessarily positive, the difference $F^+-F^-$
here can have either sign, depending on the choice of the
interpolation functions. Without the antitrapping current
($a(\phi)\equiv 0$), the condition $F^+=F^-$ that makes
this correction vanish is identical to Eq.~(\ref{MU}) in
the introduction.

\paragraph*{Choice of functions:}
In order to make all three corrections cited above vanish,
we need to satisfy simultaneously three conditions, namely,
\begin{equation}
F^+ = F^- \quad H^+ = H^- \quad Q^+ = Q^-.
\end{equation}
For fixed double-well and tilting functions $f$ and $g$, we
have at our disposal three interpolation functions: the
diffusivity $q(\phi)$, the source function $h(\phi)$ and
the antitrapping current profile $a(\phi)$. The new element
here is the antitrapping current. If it is absent, only
two interpolation functions are available. It is then, of course,
easy to satisfy two out of the three conditions. For example,
choosing $h$ odd in $\phi$ and $q(\phi)=1-q(-\phi)$, respectively,
will automatically satisfy the interface stretching and
surface diffusion conditions. However, as already discussed
in the introduction and also by Almgren for a thermal 
model \cite{Alm}, all three of them can be satisfied 
only for a weak contrast in the bulk diffusivities, which
of course excludes the one-sided case of interest here.
The problem is that, in order to satisfy the integral 
conditions shown above, the interpolation functions need
to be non-monotonous or even to change sign, which leads
to strong higher-order correction terms or even to the 
emergence of singularities.

It is interesting to note here why the corrections to the Stefan
condition, namely interface stretching and surface diffusion, 
which were not computed in Ref. \cite{KarRap}, vanish for
the symmetric model of solidification, $q(\phi)=1$, 
$a(\phi)\equiv 0$. Obviously, surface diffusion does just not
arise for a constant $q(\phi)$. But, furthermore, $p(\phi)$ reduces
to $h(\phi)-1$, and therefore the two conditions $F^+=F^-$ and
$H^+=H^-$ become identical, such that the ``miraculous'' choice 
of $h(\phi)$ odd in $\phi$ which ensured $F^+=F^-$ in 
Ref. \cite{KarRap} also cancels the interface stretching correction.

The more involved one-sided case is cured with the help
of the antitrapping current,
which offers an additional degree of freedom to
satisfy the third condition. The only place where the function
$a(\phi)$ appears in the final results of the matched asymptotics
is in the first-order concentration profile $U_1$, and more
precisely in the function 
$p(\phi)=[h(\phi)-1-2a(\phi)\partial_\eta\phi_0]/q(\phi)$.
A suitable choice for the function $a(\phi_0)$ can be obtained
by a simple analogy with the symmetric model of solidification. 
For the standard choices $a(\phi)\equiv 0$ and $h(\phi)=\phi$, 
we have $p(\phi)=\phi-1$. {\em The same} function $p(\phi)$ can
be recovered in the one-sided case if we choose
\begin{eqnarray}
q(\phi) & = &(1-\phi)/2, \\
h(\phi) & = &\phi, \\
a(\phi) & = & \frac{1}{2\sqrt{2}},
\end{eqnarray}
since we can exploit the fact that 
$\partial_\eta \phi_0 = -(1/\sqrt{2})(1-\phi_0^2)$. Then,
all the solvability integrals are identical to those
calculated in Ref.~\cite{KarRap} for the symmetric model
in the isothermal variational formulation.

Essentially, this ``trick'' solves the problem because 
it makes the two conditions $F^+=F^-$ and $H^+=H^-$ 
identical, as for the symmetrical model. 
The same strategy can be applied to obtain other possible
phase-field formulations. For any ``source function'' $h(\phi)$
and diffusivity $q(\phi)$, the equivalence to the analogous 
symmetric model can be obtained by requiring $p(\phi)=h(\phi)-1$,
which yields
\be
a(\phi)=\frac{[h(\phi)-1][1-q(\phi)]}{\sqrt{2}(\phi^2-1)}.
\ee 
For example, the function $U_1$ of the symmetric model in the 
variational formulation of Ref.~\cite{KarRap}, which uses 
$h(\phi)=\tilde g(\phi)= 15(\phi-2\phi^3/3+\phi^5/5)/8$,
can be recovered for $q(\phi)=(1-\phi)/2$ by
\be
a(\phi)=\frac{[(3\phi^3-7\phi)(\phi+1)]/8 +1}{2\sqrt{2}}\,.
\ee
Since this model is known to be less efficient, we have 
not investigated further this alternative.

\paragraph*{Kinetic coefficient:}
For low-speed solidification, kinetic effects are usually
negligibly small, and therefore we want to make the kinetic
coefficient vanish. This is possible because it consists of
two contributions of opposite signs. Converting 
Eq.~(\ref{eq:beta}) back to dimensional units, we find
(in the following, we will assume $F^+=F^-\equiv F$)
\begin{equation}
\beta = a_1\frac{\tau}{\lambda W} \left\{ 1 - a_2 \frac{\lambda W^2}{\tau D}
             \left[1 + (1-k) U_0\right]\right\},
\label{eq:betadim}
\end{equation}
\begin{equation}
 a_2 = \frac{K+JF}{2I}.
\end{equation}
For $k=1$ (constant concentration jump), this is identical
to the expression for the symmetric model, and $\beta=0$
can be achieved by choosing $\lambda = (\tau D)/(a_2 W^2)$.
For $k\neq 1$, the kinetic coefficient depends on $U_0$,
the average value of $U$ in the diffuse interface. The
physical meaning of this dependence can be understood as
follows. The second term in the expression for $\beta$ 
arises from the additional driving force supplied to
the interface by the redistribution of solute 
inside the diffuse interface. For $k\neq 1$,
the amplitude of this redistribution depends on the local
state of the interface, since the concentration jump depends
on temperature, curvature, and kinetics. To see this, recall
that $U_0 = -z_i/l_T - d_0 {\cal K} - \beta_0 V_n$, where
$\beta_0 = a_1\tau/(\lambda W)$, according to the dimensional 
version of Eq.~(\ref{a1obtain}), and furthermore that 
$c_l/c_l^0=1+(1-k)U$, and the concentration 
jump from solid to liquid is $c_l(1-k)$. 

As a consequence, the interface kinetics depends on the 
local geometry and velocity of the interface, and it is
not possible to make $\beta$ completely vanish by the
same choice as before. Among the correction terms,
$d_0{\cal K}$ and $\beta_0 V_n$ are usually small, but
no general statement can be made about the magnitude
of $z_i/l_T$. Two strategies are possible to tackle this 
problem. The first is to choose a temperature-dependent 
phase-field relaxation time,
\begin{equation}
\tau = \tau_0 [1 - (1-k) z/l_T].
\label{eq:tautemp}
\end{equation}
This does not change the asymptotic analysis for the
$\phi$ equation since the $z$-dependent part does not
contribute to the solvability conditions. It is sufficient
to replace $\tau$ by Eq.~(\ref{eq:tautemp}) in 
Eq.~(\ref{eq:betadim}). With the usual choice 
$\lambda = (\tau_0 D)/(a_2 W^2)$, the residual kinetic
coefficient is
\begin{equation}
\beta = \beta_0 (1-k) (d_0 {\cal K} + \beta_0 V_n),
\end{equation}
with $\beta_0 = a_1 \tau_0 / (\lambda W)$. The 
temperature-dependence is eliminated, but curvature and 
velocity corrections to $\beta$ remain.

The second strategy is to introduce a $U$-dependent
phase-field relaxation time,
\begin{equation}
\tau = \tau_0 \left[1 + (1-k)U\right].
\end{equation}
The idea is to make both terms of Eq.~(\ref{eq:beta}) 
contain the same prefactor $\left[1+(1-k)U_0\right]$ such that the
compensation of the two terms is independent of $U_0$.
This time, the solvability conditions for $\phi_1$ and $\phi_2$ 
are modified. The former yields a the new expression for $U_0$
\be
U_0= \frac{-\nu z_i - \alpha v_n - \kappa}{1+\alpha v_n (1-k)} .
\ee
Equation~(\ref{eq:solvability2}) that yields $\bar U_1$ becomes
\begin{eqnarray}
 & &
a_1\left\{\frac{v_n}{2} \left[1+(1-k)U_0\right] K - J \bar U_1\right\} \nonumber 
\\
 & & \qquad \mbox{}
-\alpha v_n (1-k) \left[1+(1-k)U_0\right] 
   \left\{I\bar U_1-\frac{v_n}{2} \left[1+(1-k)U_0\right] 
K'\right\} = 0,
\end{eqnarray}
where the new solvability integral,
\begin{equation}
K' = - \int_{-\infty}^\infty \,d\eta\,(\partial_\eta \phi_0)^2
    \int_0^\eta p(\phi_0(\xi)) \,d\xi ,
\end{equation}
equals $K'=0.1869$ for the choice of interpolation functions 
given above. A straightforward calculation yields
\begin{equation}
\bar U_1 = \frac{v_n}{2} \left[1+(1-k)U_0\right] \frac{K}{J}
 \left\{ \frac{1+\alpha v_n(1-k)\left[1+(1-k)U_0\right][K'J/(KI)]}
       {1+\alpha v_n (1-k) \left[1+(1-k)U_0\right]}\right\}.
\end{equation}
An expansion of this result in $v_n$ shows that the leading 
order prefactors of the two terms in $\beta$ originating from $U_0$ 
and $U_1$ are indeed the same. Furthermore, it can be seen that 
all higher order corrections are proportional to
$\alpha v_n(1-k) = \beta_0 V_n(1-k) = [a_1\tau_0/(W\lambda)]V_n (1-k)$.
As long as this quantity is much smaller than unity, the
resulting residual kinetics should be small.

\subsection{Limits of validity and expansion parameters}
In the numerical calculations presented below, we 
obtain converged quantitative results for values 
of $\epsilon=W/d_0$ much larger than unity, even
though we have used $\epsilon$ as a small 
expansion parameter in the thin interface analysis. 
This raises the question: what is the domain of validity of this 
expansion ? A rigorous answer to 
this question would in principle require to carry
out the expansion in $\epsilon$ at one more order to
determine when the additional corrections to the boundary
conditions are negligible for a given set of growth conditions.
This represents a formidable analytical task that is beyond the scope
of this work. We can, however, use
dimensional arguments to place bounds on the validity of the
thin interface analysis. We shall conclude from the foregoing arguments
that $\epsilon$ need not be small for this analysis to be
valid, consistent with the numerical findings; $W$ only needs
to be smaller than a characteristic length $l_c\gg d_0$, which depends
generally on the growth conditions.

The expansion defined by Eqs.~(\ref{formalexpansion1})
and (\ref{formalexpansion2}) assumes that $\epsilon$ is small
and that the functions $\phi_n$ and $U_n$ are  
of order unity. The magnitudes of the
functions $U_n$, however, are not known 
without specifying the outer problem. For typical
growth conditions, the variation of concentration along
the interface due to capillarity and interface kinetics
is small. In particular, the velocity-dependent form of the
Gibbs-Thomson condition implies that
$|U+\nu z| \ll 1$ in the diffuse interface region as long as
$\kappa + \tilde \beta v_n \ll 1$, and that therefore 
the right hand side of Eq.~(\ref{dimless1}) contains 
small terms other than $\epsilon$. To define a 
diffusion field that is of order unity in the interface
region, consistent with the choice of $\epsilon$ as a 
small expansion parameter, one would need to rescale 
the combination $U+\nu z$ inside
the interface by some characteristic mean interfacial value 
of the diffusion field, $\bar U$,
which depends on the outer solution.
This procedure, however, does not
change the results of the asymptotic
analysis because it amounts to
a simple change of variable. For convenience, 
we have therefore opted to keep the expansion parameter $\epsilon$,
which is independent of the outer solution.
It is clear from the above arguments, however, 
that this expansion is valid as long as
\be 
\epsilon \bar U\ll 1. \label{cv1}
\ee
Since $\bar U$ is typically small, $\epsilon$ need not be
small for the expansion to be valid.

To make Eq. (\ref{cv1}) more transparent,
it is useful to re-express this constraint
in terms of the interface velocity $V_n$ 
and the local radius of curvature $R$. 
Up to coefficients of order unity, which we
do not consider, and assuming that the
velocity is positive, it follows dimensionally that
$\bar U\sim |U + \nu z| \sim d_0/R + \beta V_n + WD/V_n$,
where $d_0/R$ and $\beta V_n$ are capillary and kinetic
corrections originating from the velocity-dependent form of the
Gibbs-Thomson condition, and $WD/V_n$ originates from 
solute diffusion in the diffuse interface region.  
 The product $\epsilon \bar U$ is therefore of order
$W/R + \beta V_n W/d_0 + W^2 V_n / (d_0 D)$. In terms
of $\epsilon$, the dimensionless kinetic coefficient
$\tilde \beta$, and the Peclet number $p$,  
Eq. (\ref{cv1}) can be rewritten as
\be
W/R + p(\tilde \beta + \epsilon) \ll 1.\label{cv2}
\ee
The same estimation can be obtained directly from the
expressions for $U_0$ and $U_1$ calculated above.
Convergence is hence limited by two independent conditions,
linked to the local curvature and velocity, respectively.
The first condition, $W/R \ll 1$, states that the interface 
thickness must be much smaller than the local radius of 
curvature. The interpretation of the second condition, 
$p(\tilde \beta + \epsilon) \ll 1$, depends on the
physical value of the kinetic coefficient to be simulated. 
In the present work, we focus on the limit of
vanishing kinetic effects relevant for small
growth velocity ($\tilde \beta = 0$), which is achieved  
by setting $\tau \sim \lambda W^2/D$. Therefore,
the limiting condition is $p\epsilon \ll 1$,
which can also be rewritten as
$\tau V_n/W \ll 1$. In practice, we found that the 
convergence starts to break down for $\tau V_n/W \sim 0.2$ 
or $W/R\sim 0.2$, although occasionally slightly larger values of
$\tau V_n/W$ could be used. 
 
Defining the diffusion length $l=D/V_n$,
Eq. (\ref{cv2}) can also be rewritten in the form
$W/\ell_c \ll 1$, where  
$\ell_c \equiv d_0 / (d_0/R + \beta V_n + W/l)$.
This shows that the true small
parameter $\epsilon \bar U$ can always be expressed
as the ratio of $W$ and some characteristic scale
scale $\ell_c$ which is much larger than $d_0$ and
which depends on experimental growth conditions.

Finally, it is in principle possible
to use the interface Peclet number
$p=WV/D$ as a small expansion parameter
in the thin-interface analysis, as for
the solidification of pure melts with symmetrical
diffusion \cite{KarRap}. However, this 
choice is not optimal for the case of asymmetrical diffusion
considered here for technical reasons. 
In particular, the interface stretching and surface
diffusion terms appear at second order and third order,
respectively, in an expansion in $p$. In contrast,
they appear both at second order in the $\epsilon$
expansion. Therefore, the latter is preferable for
clarity of exposition, with the caveat that it is
necessary to consider the outer region 
to obtain the true condition of validity of
this expansion expressed by Eqs. (\ref{cv1})-(\ref{cv2}),
or equivalently by the condition $W/\ell_c\ll 1$.

\subsection{Anisotropy}
To include anisotropy, it is sufficient to proceed in the standard
manner, that is, make $W$ and $\tau$ orientation-dependent, as
in Refs.~\cite{KarRap,KarmaPRL}, 
\begin{equation}
\label{anisotropicW}
W({\mathbf{n}}) = W a_s({\mathbf{n}}) = 
  W (1-3\epsilon_4)\left[1+\frac{4\epsilon_4}{1-3\epsilon_4}
    \frac{(\partial_x \phi)^4 + (\partial_y\phi)^4}{|\nabla\phi|^4}\right],
\end{equation}
\begin{equation}
\tau({\mathbf{n}}) = \tau_0 a_s^2({\mathbf{n}}).
\end{equation}
Here, it is understood that $\tau_0$ might be replaced by
its temperature- or $U$-dependent version. As a consequence, 
the standard result for the anisotropic capillary length is 
recovered. For the interface kinetics, the orientation dependence
appears together with $\tau_0$ in all the above results. 
Finally, note that the interface thickness also appears 
as a prefactor in the antitrapping current. However, since 
the anisotropy of $W$ itself is small (recall that the anisotropy
of the capillary length is $15$ times larger that the one of $W$
for fourfold symmetry), only a small error will be made if the 
actual orientation-dependent interface width is replaced by its
mean value in this term.

\section{Numerical tests}
\label{simuls}

We have simulated the phase-field model
of the directional solidification of a 
dilute binary alloy defined by the anisotropic version
of Eqs. (\ref{fin1}) and (\ref{fin2}) for parameters 
corresponding to the impure succinonitrile (SCN) alloy 
of Ref. \cite{Pocheau}. The alloy parameters together
with the values of the pulling speed and the
temperature gradient are listed in
Table \ref{table.param}. The chosen pulling speed is ten 
times the value for the onset of the Mullins-Sekerka instability. 
For these parameters, the capillary length is several orders
of magnitude smaller than the thermal length or the
diffusion length. Since typical cell widths are $\sim 100~ \mu$m
or $\sim 10^4 d_0$ and computations are only feasible if
one cell width $\sim 10^2$ grid points,
we are forced to use values of $W$ much larger than $d_0$, typically 
$W/d_0 \simeq 10$ to $100$. We will see that, 
with the present phase field model, it is possible
to obtain well converged results even with such large $W/d_0$ ratios.

\begin{table}
\begin{tabular}{|l|c|}
\hline
$|m| c_{\infty}$ (shift in melting temperature) & $2$ K \\ \hline
$D$ (diffusion coefficient) & $10^{-9}$ m$^2$/s \\ \hline
$\Gamma$ (Gibbs-Thompson coefficient) & $6.48 \times 10^{-8}$ K\,m \\ \hline
$V_p$ (pulling speed) & 32 $\mu$m/s \\ \hline
$G$ (thermal gradient) & 140 K/cm \\ \hline
$d_0$ (capillary length)  &  $1.3 \times 10^{-2}$ $\mu$m \\ \hline
$l_T$ (thermal length)  & $3.33 \times 10^2$ $\mu$m \\ \hline
$l_D$ (diffusion length) & 60 $\mu$m  \\ \hline
$k$ (partition coefficient) & $0.3$ \\ \hline
\end{tabular}
\caption{Parameters for
the impure succinonitrile (SCN) alloy system
of Ref. \cite{Pocheau} used in the phase-field
simulations and corresponding 
characteristic length scales
for directional solidification. 
The anisotropy of the interfacial free energy
is taken to be $\epsilon_4=0.007$ (0.7\% anisotropy).} 
\label{table.param}
\end{table}
  
To choose the phase field model parameters, we first
note that the ratio of the capillary and thermal lengths, 
$\nu=d_0/l_T = 4\times 10^{-5}$, and the dimensionless pulling speed 
$v_p = V_p d_0/D=4.16 \times 10^{-4}$ completely specify the
interface evolution in the sharp-interface equations. 
This can be seen by scaling length and time in
these equations by $d_0$ and $d_0^2/D$, respectively. 
In the phase-field model, we have the additional
length $W$ and converged results
should be independent of the ratio $\epsilon=W/d_0$.
Note that for anisotropic surface tension 
$W(\mathbf{n})=W a_s(\mathbf{n})$
with $a_s(\mathbf{n})$ given by Eq.~(\ref{anisotropicW}).
In a given simulation, we fix $\epsilon=W/d_0$ and
hence $\lambda=a_1 \epsilon$ from Eq. (\ref{eqd0}). 
Furthermore, we use a temperature- and orientation-dependent
relaxation time $\tau$ as specified in the previous section
together with the relation $\tau_0 =a_2 \lambda W^2/D$,
which makes the interface kinetic coefficient vanish for
all temperatures and orientations, and we scale 
lengths by $W$ and time by $\tau_0$ in the phase-field
equations. The scaled phase-field equations then only depend
on $\epsilon$ through the dimensionless parameters 
$\tilde D=D\tau_0/W^2=a_1 a_2 \epsilon$, 
$\tilde V_p=V_p\tau_0/W=v_p a_1 a_2 \epsilon^2$,
and $\tilde l_T=l_T/W=1/(\epsilon \nu)$. Writing
out explicitly all the interpolation functions,
and taking into account the contributions of the
anisotropic $W(\mathbf{n})$ in the functional 
derivative, the equations read
\begin{eqnarray}
\left[1 - (1-k)\frac{z-\tilde V_p t}{\tilde l_T}\right]{a_s(\mathbf{n})}^2
\frac{\partial \phi}{\partial t} &=& 
   \vec\nabla\left[a_s(\mathbf{n})^2 \vec\nabla \phi\right]  \nonumber \\
& & \mbox{}
  +\partial_x \left(\left|\vec\nabla \phi\right|^2 a_s(\mathbf{n}) 
      \frac{\partial a_s(\mathbf{n})}{\partial(\partial_x\phi)}\right)
  +\partial_y \left(\left|\vec\nabla \phi\right|^2 a_s(\mathbf{n}) 
      \frac{\partial a_s(\mathbf{n})}{\partial(\partial_y\phi)}\right) 
	  \nonumber \\
& & \mbox{}
+\phi-\phi^3-\lambda\, \left(1-\phi^2\right)^2 
   \left(U+ \frac{z-\tilde V_p t}{\tilde l_T}\right),\label{pfn1}\\
\left(\frac{1+k}{2} - \frac{1-k}{2} \phi \right) 
   \frac{\partial U}{\partial t}&=& 
\vec \nabla \cdot \left( \tilde D \frac{1-\phi}{2}
\,\vec\nabla U + \frac{1}{2\sqrt{2}} \left[1+(1-k)U\right] 
\frac{\partial \phi}{\partial t}\,\frac{\vec \nabla \phi}
{|\vec \nabla \phi|}\right)\nonumber\\
& &\mbox{} + \left[1+(1-k)U\right] 
   \frac{1}{2}\frac{\partial \phi}{\partial t}, \label{pfn2}
\end{eqnarray}
where $x$ and $z$ are in units of $W$ and 
$t$ is in unit of $\tau_0$.
Simulations are repeated with different
values of $\epsilon$ to study the convergence.    
The equations are discretized
on a square lattice; some details are given in the appendix. We have 
used a grid spacing $\Delta x /W=0.8$ in most of the simulations,
but also used a finer resolution $\Delta x /W=0.4$ to study
the effect of the discretization.
For the time evolution, we have used an explicit Euler scheme with
a time step chosen below the threshold of numerical instability
for the diffusion equation in two dimensions, $\Delta t < (\Delta x)^2/(4D)$. 

\subsection{Stability spectrum}

\begin{figure}[ht]
\includegraphics[width=10cm]{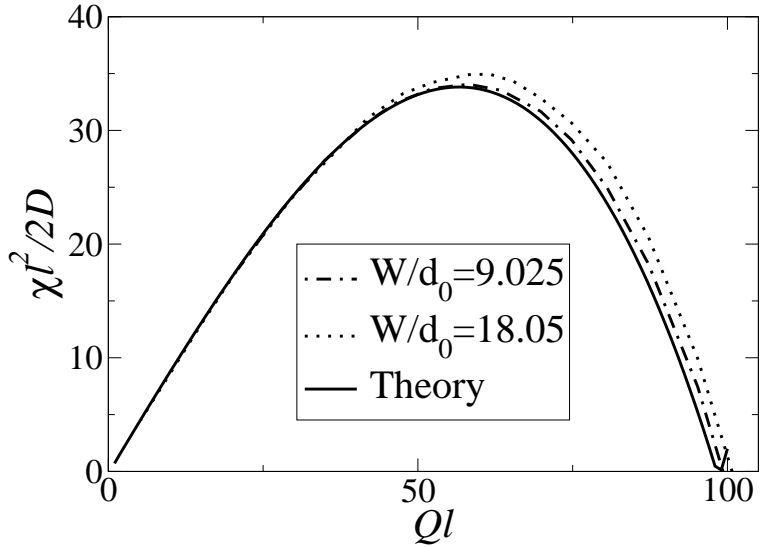}
\caption{Comparison between the linear stability spectrum of a planar
steady-state interface computed with the phase-field model for different
interface thicknesses (dot-dashed and dotted lines) and the 
Mullins-Sekerka theory \protect\cite{MS} (solid line).
Here, $\chi$ is the growth rate of a sinusoidal perturbation 
of wave number $Q$, and $l=2D/V_p$
is the diffusion length. The parameters are for an impure SCN alloy
system described in the text with $V_p=32$ $\mu$m/sec and $G=140$ K/cm.}
\label{figMS}
\end{figure}

\begin{figure}[ht]
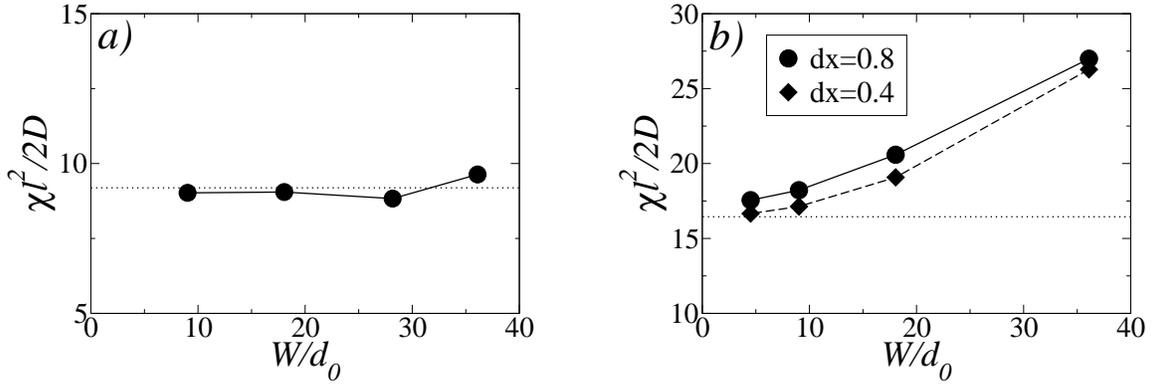

\vspace{0.5cm}
\includegraphics[width=7cm]{fig3a.eps}\hspace{1cm}
\includegraphics[width=7cm]{fig3b.eps}
\caption{Convergence of the growth rate 
$\chi(Q)$ as a function of $W/d_0$ for:
a) $Ql=10.5$, and b) $Ql=87.3$. The dotted lines are the predictions
of the Mullins-Sekerka analysis.}
\label{figsigk}
\end{figure}

We have numerically calculated the stability spectrum of a planar 
steady-state interface. To this end, the system was initialized
with a planar interface at its steady-state position. The
concentration in the liquid was set to the exponential
steady-state solution of the free boundary problem. A small
sinusoidal perturbation of amplitude $A\ll W$ and wave number $Q$ was 
then applied, and its time evolution was followed by extracting 
successive interface positions. It follows an exponential increase 
or decay, and the growth rate $\chi(Q)$
was extracted by a fit of the perturbation amplitude versus time.

In Fig. \ref{figMS}, we compare the results from the numerical 
simulations to the analytical solution for the Mullins-Sekerka 
stability spectrum of the free-boundary problem of
Eqs.~(\ref{fbnodim1})--(\ref{fbnodim3}). 
The convergence is better for smaller wave numbers, 
which is perfectly reasonable since the ratio of perturbation
wavelength to interface thickness scales with $Q$. For $W/d_0=9.025$, 
the phase field model gives a good 
agreement for almost the whole range of wave numbers, including the
maximum, which is the most important part of the spectrum.
In Fig.~\ref{figsigk}, we plot the growth rate $\chi(Q)$
of two selected modes versus the ratio $W/d_0$, which shows a fast
convergence. For $\Delta x/W=0.4$, the results are fully
converged to the theoretical value for $W/d_0=4.51$
even for the mode with high wavenumber. It can
also be seen that the larger grid spacing of $\Delta x/W=0.8$
introduces slight corrections that are due to the lattice
pinning effect (see Ref.~\cite{KarRap}).

\begin{figure}[ht]
\vspace{0.5cm}
\includegraphics[width=8cm]{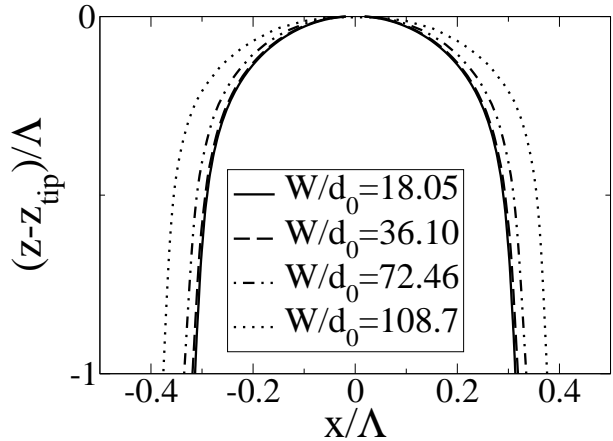}\hspace{0.5cm}
\vspace{0.5cm}
\caption{Convergence of the shape of steady-state deep cells as a 
function of interface thickness. Lengths are scaled by the
cell spacing $\Lambda=22.5$ $\mu$m, $V_p=32$ $\mu$m/sec and
$G=140$ K/cm.}
\label{figcells}
\end{figure}

\subsection{Cell shapes}

To asses the convergence of the models in the nonlinear regime, 
we have computed shapes of steady-state cells for various values 
of $W/d_0$. The simulation box contains half of a cell, with 
no-flux boundary conditions along the cell center and the groove.
We have considered narrow cells of spacing 
$\Lambda=1732.8 \times d_0=22.5$ $\mu$m, since we want to test the
convergence of the model for small tip radii; in an extended
system, these cells would be unstable to a cell-elimination 
instability that leads to a doubling of the cell spacing. As initial
condition we set $c_l=c_\infty$, $c_s=k c_l$ (which, with the
definition of Eq.~(\ref{udefsharp}) and using $c_l^0=c_\infty/k$, corresponds
to $U\equiv -1$ in the whole system), and add a small sinusoidal
perturbation to the interface, with a wavelength equal to
the cell spacing and its maximum located on the boundary. 
After a transient where the interface recoils,
it reaches steady state in the form of a half cell.   

\begin{figure}[ht]
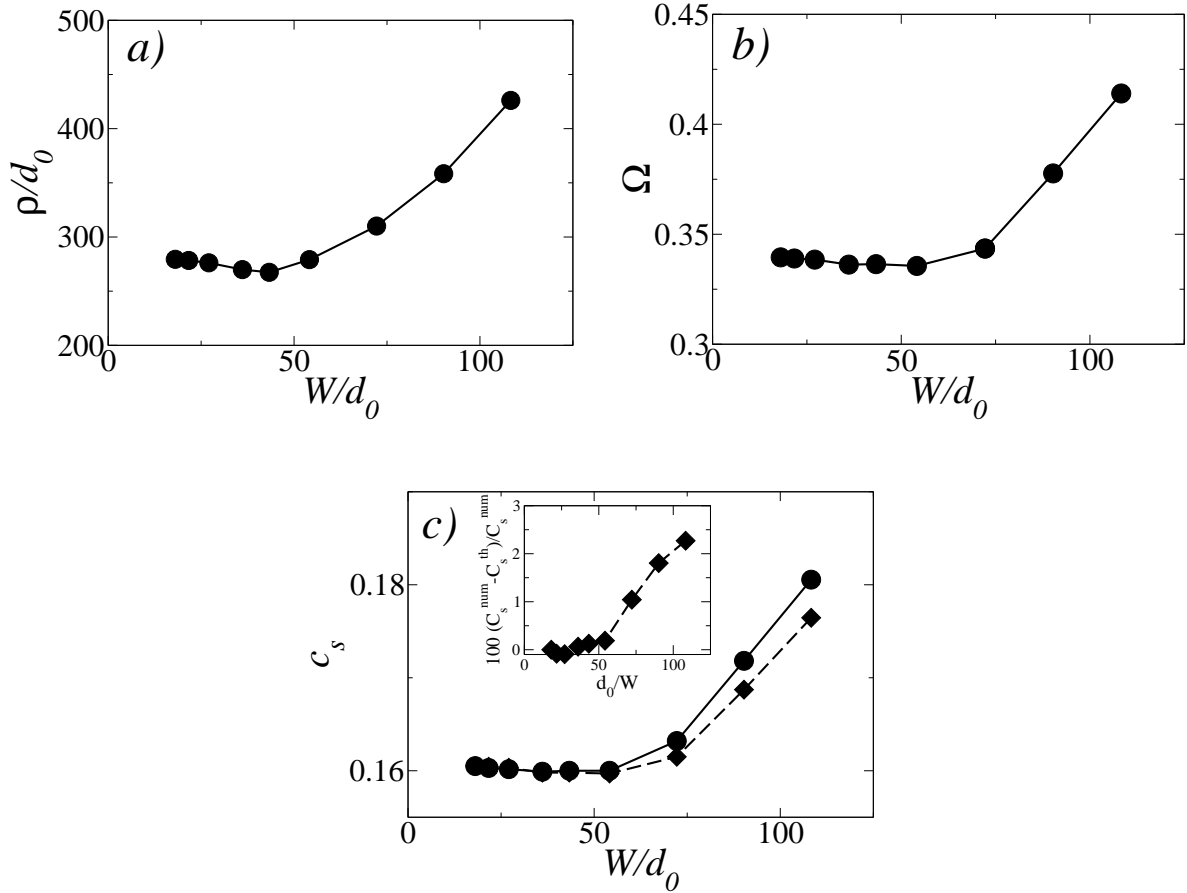

\includegraphics[width=7.5cm]{fig5a.eps}\hspace{0.5cm}
\includegraphics[width=7.5cm]{fig5b.eps}\\[0.9cm]
\includegraphics[width=7.5cm]{fig5c.eps}
\vspace{0.5cm}
\caption{Convergence as a function of 
interface thickness of various quantities associated with  
steady-state cell shapes: a) tip radius $\rho$, b) 
dimensionless tip undercooling $\Omega$ and 
c) solid concentration in the center of the cell.  
The diamonds (dashed line) in c) correspond to the values
calculated from the Gibbs-Thompson condition using the tip radius of the
phase-field shape. The inset shows the relative
error of the phase-field results with 
respect to the Gibbs-Thomson prediction.} 
\label{figconv}
\end{figure}

The resulting shapes are shown in Fig.~\ref{figcells}. For the cell shapes the
convergence is faster than for the growth rate, and already simulations with
$W/d_0 \simeq 50$ are well converged. To show more 
clearly the difference in the speed
of convergence, we plot in Fig.~\ref{figconv} the tip radius
$\rho/d_0$, the tip undercooling $\Omega=1-z_{tip}/l_T$ (where
$z=0$ corresponds to the position of the steady-state interface), 
and the solute concentration in the solid in the center of
the cell. For the latter, we compare the values that are directly 
obtained from the simulations (that is, the value of the field $U$ in 
the center of the cell) to the value expected from the Gibbs-Thomson
condition and partition relation at the interface,
$c^{th}_s=k[k+(1-k)(\Omega-d_0/\rho)]$, where the values of 
$\Omega$ and $\rho$ are obtained from the numerical 
results (Figs.~\ref{figconv}a-b.) Again, all the quantities are well
converged for $W/d_0 \simeq 50$, and even for
the ratio $W/d_0=72.2$ (corresponding to $\rho/W \simeq
4$), the error in the tip radius for the phase field model is only
about 15\%, while the equilibrium solute concentration condition at
the interface is satisfied within an error of about 1\%. The error in this
latter condition is small, even for the largest values of $W/d_0$ used
(Fig.~\ref{figconv}c).  
Since microsegregation
is important for metallurgy, the precise calculation of the
solute concentration in the solid is an important new 
feature of the present model.

\section{Conclusions and perspectives}
\label{conclusions}

We have presented a detailed asymptotic analysis of the phase-field
model for alloy solidification that was introduced in Ref.~\cite{KarmaPRL},
and we have simulated directional solidification of a dilute binary
alloy. We have found a very good quantitative agreement
with the Mullins-Sekerka stability spectrum of a planar
interface for typical experimental control parameters. For 
solidification cells, we found that the solute concentration
inside the solid agrees self-consistently with the prediction
of the Gibbs-Thomson condition, in contrast to earlier models where
the microsegregation was only qualitatively 
reproduced \cite{Warren95}.

This advance relies on a solution of the complete problem of
canceling all relevant thin-interface corrections to the original free 
boundary problem. This opens the way for quantitative comparisons between 
experiments and simulations both in two and three dimensions, 
with the concomitant possibility of testing the 
theories and concepts used to interpret microstructural
pattern formation, as was previously done for dendritic solidification.

The present work can be extended along several lines. For example, it 
has been demonstrated that the concept of the antitrapping current can 
be generalized to two-phase solidification, which makes it possible
to study eutectic or peritectic composite growth with excellent
precision \cite{Folch03}. Also, the present one-sided model can be combined
with a symmetric thermal model to yield a quantitative thermosolutal
model of solidification \cite{Beckermann}. A small solute 
diffusivity in the solid can also be introduced without appreciable 
modifications of the present analysis. Finally, the antitrapping
current, which was used here to restore the equilibrium partition
relation, can also be used to obtain a non-vanishing, specified 
trapping. This is especially important to extend this model to
the whole range of solidification velocity relevant for experiments.
In addition, the present model should be applicable to model Hele-Shaw 
flows when the viscosity of one fluid is much smaller than that of the other.

From a broader perspective, 
this progress revives the hope
of using the phase-field method as an efficient and 
fully predictive tool for other
free boundary and interface growth problems where 
the dynamics of the two media are not necessarily symmetric, even 
outside the framework of systems described by a Lyapounov
functional. A key element of this progress is the use of
non-variational terms which provide additional freedom to
obtain the correct mapping between a diffuse interface model and
a desired free-boundary problem, such as the antitrapping current here,
and other terms in other contexts \cite{Folchetal}. 
It is important to emphasize that the interface is spatially diffuse
and all interpolation functions are smooth in the 
present phase-field model. Hence, this model remains simple to 
implement numerically in comparison to other methods
that combine sharp and diffuse interface ingredients \cite{Gold,Amberg}.

Let us conclude with a few remarks on the formulation of the model
itself. The thermodynamic derivation presented here, which is an
alternative to previous expositions of the same model \cite{encyclo},
establishes new connections to other phase-field models of alloy 
solidification. As mentioned before, early phase-field models of
alloy solidification were plagued by a dependence of the surface
tension on the interface thickness that arose from the coupling
between the phase-field and concentration equations \cite{Wheeler,Caginalp93}. 
This problem was solved later by the introduction of two separate concentration
fields, one for the solid and one for the liquid, and by interpreting
the interface as a mixture of two phases \cite{Steinbach98}. 
The requirement of local equilibrium between the two phases 
then allows to eliminate one of the concentration fields \cite{Kim99}. 
The resulting model has a surface tension that is independent 
of the interface thickness and can be used for arbitrary 
phase diagrams; however, some thin-interface effects remain,
in particular surface diffusion \cite{Kim99}. 

In our derivation, we have succeeded in constructing a quantitative
model for an ideal dilute binary alloy with a single 
concentration field, but two different interpolation
functions of the phase field for entropy and internal energy 
density. This is appealing from a thermodynamic viewpoint,
since it maintains the interpretation of the concentration 
as a local quantity rather than a two-phase mixture.
An interesting task would be to generalize this approach 
to arbitrary phase diagrams and multi-phase solidification.

\acknowledgments
We thank Hermann-Josef Diepers for many interesting discussions.
R. F. was supported by the European Community through a Marie-Curie
fellowship. This research is supported by NASA and U.S. DOE through Grant 
No. DE-FG02-92ER45471 and funds from the Computational Materials
Science Network. B. E. acknowledges financial support from the Ministerio de
Ciencia y Tecnolog\'{\i}a (Spain).  

\appendix

\section{Anomalous interface kinetics}
We give here a more detailed discussion of the interface 
kinetics in a phase-field model without antitrapping 
current, and with source and diffusion functions
given by $h(\phi)=\phi$ and $q(\phi)=(1-\phi)/2$. We will
see that in this model, the interface kinetics has 
logarithmic corrections. This occurs whenever in the
limit $\phi\to 1$ the ratio $(h-1)/q$ does not vanish
(i.e., remains finite or diverges). Note that, in physical
terms, the two functions describe the thermodynamic driving
force for solute redistribution during the phase transformation
and the diffusivity, respectively. If the latter vanishes
faster than the former, the redistribution cannot be
completely accomplished on the solid side of the interface,
and trapping occurs. We will now analyze this effect in more
detail.

Our starting point is Eq.~(\ref{eq:intdiffusionin1}) for
the first-order diffusion field in the inner region.
Without antitrapping current, its solution is
\begin{equation}
U_1 = {\bar U}_1 + 
  \frac{v_n}{2} \left[1+(1-k)U_0\right] \int_0^\eta 
  \frac{\phi_0(\xi)-1}{q[\phi_0(\xi)]} \,d\xi,
\end{equation}
which for the above choice of functions becomes
\begin{equation}
\label{eq:naivesol}
U_1 = {\bar U}_1 - v_n \left[1+(1-k)U_0\right] \eta  \, .
\end{equation}
This solution, however, is not appropriate since it
cannot be matched to the outer solution in the solid,
which for a steady state is just a constant. The
problem is that we have neglected terms in 
Eq.~(\ref{eq:intdiffusionin1}) that, for this solution,
would not be small, which makes the calculation 
inconsistent. To see this, it is sufficient to remark
that both the diffusion term (proportional to $q$) and the 
redistribution term (proportional to $h-1$) become 
exponentially small inside the solid. In contrast, for the 
above solution, the time derivative of $U_1$ (equivalent 
to $v_n\partial_\eta$ in the moving frame) gives a term of order
$\epsilon$ in the equation for $U$, and hence becomes larger
than the two mentioned terms far enough in the solid, which violates 
the counting of orders. In order to get a solution valid everywhere 
inside the solid, this term has to be included
in the equation for $U_1$ which becomes
\begin{equation}
\label{eq:newdiff}
\partial_\eta [q(\phi_0)\partial_\eta U_1] = 
  \frac{v_n}{2} \left[1+(1-k)U_0\right] \partial_\eta \phi_0 
  - \epsilon v_n \left(\frac{1+k}{2}-\frac{1-k}{2}\phi_0\right)\partial_\eta U_1 
\, .
\end{equation}
By integrating once and using the boundary condition of
vanishing current in the solid ($q(\phi_0)\partial_\eta U_1 \to 0$ 
for $\eta \to -\infty$), we find
\begin{equation}
\label{eq:discuss}
q(\phi_0)\partial_\eta U_1 = \frac{v_n}{2}\left[1+(1-k)U_0\right](\phi_0-1)
    -\epsilon v_n \int_{-\infty}^\eta  
            \left(\frac{1+k}{2}-\frac{1-k}{2}\phi_0(\xi)\right) 
                \partial_\xi U_1(\xi) \, d\xi \,.
\end{equation}

For the sake of simplicity, let us first discuss the case $k=1$, in
which the integral on the right-hand side is simply equal to
$U_1(\eta)-U_1(-\infty)$. It can be seen immediately that this
equation admits a solution that has the right limit, 
$\partial_\eta U_1 \to 0$ for $\eta \to -\infty$. We
proceed by constructing an approximate solution by a
matching procedure. First, remark that the left-hand side
of Eq.~(\ref{eq:discuss}) is the product of two functions
that vanish in the limit $\eta \to -\infty$. Hence, it
can be neglected in this limit, and the asymptotic
solution is
\begin{equation}
U_1(\eta) = U_1(-\infty) + \frac{1+(1-k)U_0}{2\epsilon} (\phi_0 -1) \, .
\end{equation}
In contrast, in the region of the interface, the newly
introduced term, being of order $\epsilon$, is small,
which was precisely the reason to neglect it in the
usual calculation. Therefore, in this region the solution
of Eq.~(\ref{eq:naivesol}) applies. Finally, a matching
between the two solutions is found by searching the
coordinate $\eta^\star$ where their slopes are equal,
which, using the fact that $\phi_0 = -\tanh(\eta/\sqrt{2})$,
yields
\begin{equation}
\eta^\star = - \sqrt{2} \cosh^{-1}\frac{1}{\sqrt{2\sqrt{2}\epsilon v_n}} \, .
\end{equation}
In the limit of small velocity $v_n$, this simplifies to
$\eta^\star = (1/\sqrt{2})\ln(\epsilon v_n/\sqrt{2})$.
It can be checked that, in the matching region, the two
terms (diffusion and time derivative) are of similar
magnitude, which justifies the matching procedure.

We have hence constructed an approximate solution, which
is equal to the one obtained from the standard procedure
for $\eta > \eta^\star$, and becomes a decaying exponential for
$\eta < \eta^\star$. Evaluating the solvability integrals
with this solution, we find, for example,
\begin{equation}
F^- = U_1(-\infty) - U_1(0) = 
  \frac{v_n}{2\sqrt{2}} \left[1+(1-k)U_0\right]
  \left[1-2\ln(\epsilon v_n/\sqrt{2})\right] \, .
\end{equation}
Similar terms appear also in the integral $K$. Using the
identity $\epsilon v_n = (W/d_0) (V_n d_0/D) = p$, we
find that the kinetic coefficient contains, in addition
to the usual terms linear in $p$, corrections coming
from $\epsilon F^-$ that scale as $p \ln p$. This constitutes,
for small $p$, a logarithmic correction that makes convergence
in $p$ very slow.

This calculation is an approximation, but the conclusion
that there are nonlinear correction terms is general, 
and can be easily interpreted: the anomalous kinetics
occurs because solute can escape only from a region of size 
$\eta^\star$ behind the interface, and this size scales 
logarithmically with $V_n$ (and hence $p$) in the limit of small
$p$. In this limit, the case of arbitrary $k$ can be
easily treated and yields corrections of the form 
$p\ln(k p)$. In the light of this conclusion,
a physical sense can also be given to the condition
used in the main body of the paper, namely that the
``source function'' must decay faster than the diffusivity:
under this condition, all solute can escape the advancing
interface. Note also that for a more realistic model in which 
the diffusivity becomes small but finite in the solid, the
anomalous dependence of the kinetics on $v_n$ stops for 
$v_n < q(+1)$ since then, again, all solute can escape.

\section{Discretization}
The phase-field and diffusion equations are discretized on a
square grid of spacing $\Delta x$. We use standard finite-difference
formulas, but a few details are worth mentioning.

For the Laplacian of the phase field, we use the
maximally isotropic discretization,
\begin{eqnarray}
\nabla^2 \phi_{i,j} & = & 
\frac{2}{3}(\phi_{i+1,j}+\phi_{i-1,j}+\phi_{i,j+1}+\phi_{i,j-1}) \nonumber \\
   & & + \frac{1}{3}(\phi_{i+1,j+1}+\phi_{i-1,j+1}+\phi_{i+1,j+1}+\phi_{i-1,j-1}) 
- 5\phi_{i,j}
\end{eqnarray}
which avoids the grid corrections to the anisotropy that are 
discussed in Ref.~\cite{KarRap}.

For the diffusion field $U$, we proceed by first calculating
the current in each link, and then summing up all links around
a site. On each link, the diffusion part, $j_u= - \phi \nabla U$
is calculated with the average of the phase-field according to 
$j_u = - (\phi_{i+1,j}+\phi_{i,j})(u_{i+1,j}-u_{i,j})/(2\Delta x)$
for the $x$ direction, and an analogous expression for the $y$ 
direction. The most delicate part is the antitrapping current,
$j_{at}=a(\phi)W[1+(1-k)U]\hat n \partial_t \phi$, where 
$\hat n=-\vec\nabla\phi/|\vec\nabla\phi|$ is the unit normal
vector pointing into the liquid. We first evaluate the components
of $\vec\nabla \phi$. The computation of the component parallel
to the link is straightforward. As for the component 
perpendicular to a link, for a link along the $x$ 
direction between sites $(i,j)$ and $(i+1,j)$ we use
\be
\partial_y \phi = 
\frac{\phi_{i+1,j+1}-\phi_{i+1,j-1}+\phi_{i,j+1}-\phi_{i,j-1}}{4\Delta x},
\ee
and similarly for $\partial_x\phi$ on links along $y$. From the
components of $\vec\nabla\phi$, we obtain $\hat n$. The product
$a(\phi)[1+(1-k)U]\partial_t\phi$ is then evaluated at the two end
points of the link, and its average value multiplied with the
appropriate component of $\hat n$ to obtain the current.


\begin{thebibliography}{99}


\bibitem{ARMR} L.-Q.~Chen, Annu. Rev. Mater. Res. {\bf 32}, 113 (2002);
  W.~J.~Boettinger, J.~A.~Warren, C.~Beckermann, and A.~Karma, 
  {\em ibid}. {\bf 32}, 163 (2002).

\bibitem{CrossHoh} M.~C.~Cross and P.~C.~Hohenberg,
  Rev. Mod. Phys. {\bf 65}, 851 (1993).

\bibitem{Kurz} W.~Kurz and D.~J.~Fisher, {\em Fundamentals of Solidification}
  (Trans Tech, Aedermannsdorf, Switzerland, 1992).

\bibitem{BGJ} J. Q. Broughton, G. H. Gilmer, and K. A. Jackson,
Phys. Rev. Lett. {\bf 49}, 1496 (1982).

\bibitem{MC} L. V. Mikheev and A. A. Chernov, J. Cryst. Growth {\bf 112},
591 (1991).

\bibitem{HAK} J. J. Hoyt, M. Asta, and A. Karma, Inter. Science {\bf 10},
149 (2002).

\bibitem{Wheeler} A. A. Wheeler, W. J. Boettinger, and G. B. McFadden,
  Phys. Rev. A{\bf 45}, 7424 (1992); Phys. Rev. E{\bf 47}, 1893 (1993).

\bibitem{Caginalp93} G. Caginalp and W. Xie,
  Phys. Rev. E{\bf 48}, 1897 (1993).
  
\bibitem{encyclo} A. Karma, in {\it Encyclopedia of Materials: Science and 
Technology}, edited by K. H. J. Buschow, R. W. Cahn, M. C.          
Flemings, B. Ilschner, E. J. Kramer, S. Mahajan, 
Volume 7, Elsevier, Oxford, pp. 6873-86 (2001); in
\emph{Thermodynamics, Microstructures and Plasticity}, edited by
A. Finel \emph{et al.}, Kluwer Academic Publishers, 
North Holland, pp. 65-89 (2003).

\bibitem{Steinbach98} J. Tiaden, B. Nestler, H.-J. Diepers, and I. Steinbach, 
  Physica D {\bf 115}, 73 (1998).

\bibitem{Kim99} S.~G. Kim, W.~T. Kim, and T. Suzuki, 
  Phys. Rev. E {\bf 60}, 7186 (1999).

\bibitem{Bragard}
J. Bragard, A. Karma, Y. H. Lee, M. Plapp, 
Interface Science {\bf 10}, 121 (2002).

\bibitem{KarRap} A. Karma and W.-J. Rappel, Phys. Rev. E {\bf 53}, R3017
(1996); Phys. Rev. E {\bf 57}, 4323 (1998).

\bibitem{Proetal} N. Provatas, N. Goldenfeld, and J. Dantzig,
Phys. Rev. Lett. {\bf 80}, 3308 (1998); J. Comp. Phys. {\bf 148}, 265 (1999).

\bibitem{PlaKar}
M. Plapp and A. Karma, Phys. Rev. Lett. {\bf 84},
1740 (2000); J. Comp. Phys. {\bf 165}, 592 (2000).

\bibitem{KarTip}
A. Karma, Y. H. Lee, and M. Plapp, Phys. Rev. E{\bf 61} 3996 (1999).

\bibitem{Warren95} J. A. Warren and W. J. Boettinger, 
  Acta Metall. Mater. {\bf 43}, 689 (1995). 

\bibitem{Alm} R. F. Almgren, SIAM J. Appl. Math {\bf 59}, 2086 (1999).

\bibitem{KarmaPRL} A.~Karma, Phys. Rev. Lett. {\bf 87}, 115701 (2001).

\bibitem{KurFis} W. Kurz and D. J. Fisher, {\sl Fundamentals
of solidification} (Trans Tech, Aedermannsdorf, Switzerland, 1992).

\bibitem{trapping} M.~J. Aziz, J.~Appl. Phys. {\bf 53}, 1158 (1982);
  N. A. Ahmad, A. A. Wheeler, W. J. Boettinger, and G. B. McFadden, 
  Phys. Rev. E {\bf 58}, 3436 (1998).
  
\bibitem{Eldetal01} K. R. Elder, M. Grant, N. Provatas, and
J. M. Kosterlitz, Phys. Rev. E {\bf 64}, 021604 (2001).

\bibitem{McFadden00} G. B. McFadden, A. A. Wheeler, and D. M. Anderson,
  Physica D {\bf 144}, 154 (2000).

\bibitem{MS} W. W. Mullins and R. F. Sekerka, J. Appl. Phys. {\bf 35},
444 (1964).

\bibitem{Wangetal} S-L. Wang, R.F. Sekerka, A.A. Wheeler, B.T. Murray,
S.R. Coriell, R.J. Braun, and G.B. McFadden, Physica D {\bf 69}, 189
(1993).

\bibitem{Folchetal} 
R. Folch, J. Casademunt, A. Hern\'andez-Machado, and L. Ram\'{\i}rez-Piscina,
Phys. Rev. E {\bf 60}, 1724 (1999); {\bf 60}, 1734 (1999).

\bibitem{Pocheau} M. Georgelin and A. Pocheau, 
  Phys. Rev. E {\bf 57}, 3189 (1998); Eur. Phys. J B {\bf 4}, 169 (1998).

\bibitem{Folch03} R. Folch and M. Plapp, 
  Phys. Rev. E {\bf 68}, 010602R (2003).

\bibitem{Beckermann} J. C. Ramirez, C. Beckermann, A. Karma, and H.-J. Diepers, 
  Phys. Rev. E {\bf 69}, 051607 (2004). 

\bibitem{Gold}
Y.-T.~Kim, N.~Goldenfeld, and J.~Dantzig,
Phys. Rev. E {\bf 62}, 2471 (2000).

\bibitem{Amberg}
G.~Amberg, Phys. Rev. Lett. {\bf 91}, 265505 (2003). 

\end{thebibliography}
\end{document}